\documentclass{aa}  
\usepackage{graphicx}
\usepackage{amsmath}
\usepackage{txfonts}
\usepackage{natbib}
\bibpunct{(}{)}{;}{a}{}{,} 
\usepackage[dvips]{color}
\newcommand{\realfigure}[3]{
              \begin{figure}
              \includegraphics[width=\columnwidth]{#1}
              \caption{#2}\label{#3}
              \end{figure}}

\newcommand{\referee}[1]{#1}

\newcommand{\rereferee}[1]{#1}

\newcommand{\feh}{{\rm[Fe/H]}}
\newcommand{\mh}{{\rm[M/H]}}

\def\nodata{---}

\begin{document}
   \title{Near-Infrared photometry of \referee{carbon stars in} the Sagittarius dwarf irregular galaxy and DDO 210
      \thanks{Based on data collected at the European Southern Observatory,
           La Silla, Chile, Proposal No. 71.D-0560, and archived ESO data from the 
	   Proposal No. 61.E-0273.}}
   \author{
          M. Gullieuszik
          \inst{1,2}
          \and
          M. Rejkuba\inst{3}
          \and
          M. R. Cioni\inst{4}
          \and 
          H. J. Habing \inst{5}
          \and
          E. V. Held \inst{1}    
          }

   \offprints{M. Gullieuszik}

   \institute{Osservatorio Astronomico di Padova, INAF,
              vicolo dell'Osservatorio 5, I-35122 Padova, Italy
              \\ \email{marco.gullieuszik@oapd.inaf.it,enrico.held@oapd.inaf.it}
         \and
             Dipartimento di Astronomia, Universit\`a di Padova
              vicolo dell'Osservatorio 2, I-35122 Padova, Italy
         \and
             European Southern Observatory, 
             Karl-Schwarzschild-Strasse 2,
             85748 Garching, Germany
             \\ \email{mrejkuba@eso.org}
         \and
             SUPA, School of Physics, University of Edinburgh, IfA,
             Blackford Hill, Edinburgh EH9 3HJ, UK
             \\ \email{mrc@roe.ac.uk}    
          \and
             Sterrewacht Leiden, 
             Niels Bohrweg 2, 2333 RA Leiden, The Netherlands
             \\ \email{habing@strw.leidenuniv.nl}
         }

\titlerunning{Near-Infrared photometry of the Sagittarius dwarf irregular galaxy and DDO 210}
\authorrunning{M. Gullieuszik et al.}
   \date{Received xxx; accepted xxx}
  \abstract
   {}
   {
We investigate the intermediate-age asymptotic giant branch stellar
population of two Local Group dwarf irregular galaxies to
characterize their carbon star population in near-infrared (IR).
}
   {
Our work is based on near-IR photometry complemented with optical
ground based and Hubble Space Telescope (HST) photometry. Near-IR
photometry is based on our and archival $J$ and $K_s$-band images from
SOFI near-IR array of the ESO New Technology Telescope (NTT). Optical
photometry for DDO~210 is from the EMMI optical imager of ESO NTT, while
the SagDIG optical data come from Momany et al.\
(2005).
}
   {
We show that near-IR photometry is a very powerful tool for carbon
star detection.  We recovered two out of three previously-known
carbon stars in DDO 210 and discovered six additional objects in this
galaxy which have optical and near-IR colors consistent with carbon
giants. This brings the total number of bona-fide C-star candidates in
DDO 210 to nine.  However, to confirm the nature of these objects
additional higher spatial resolution imaging or spectroscopic data are
necessary. \\ 
We detected a large population of C-star candidates in SagDIG, 18 of
which were previously identified in Demers
\& Battinelli (2002) and Cook (1987), and six new bona-fide carbon stars.
We present their optical and near-IR colors and use their luminosity
function to put constraints on the star formation history (SFH) in this
dwarf irregular galaxy.
}
   {}


\maketitle

\section{Introduction}
The Asymptotic Giant Branch (AGB) is an important, although 
short-lived phase during the final evolutionary stages of low and
intermediate mass stars.
\referee{
In unresolved galaxies with intermediate-age stellar populations AGB
stars provide a major contribution to the integrated light
\citep{renzbuzz1986}.  A good description of this evolutionary phase
is therefore fundamental to understand the integrated
properties of high redshift galaxies \citep[e.g.][]{mara+2006}.}
Its importance for the chemical evolution and
SFH in external galaxies is shaded by the fact that this is also one
of the least known evolutionary phases. Large samples of AGB stars in
systems with well defined distances, metallicities and SFHs are
necessary to constrain the theoretical models which predict their
properties.
\referee{In addition, \cite{battdeme2005} considered 
the possible standard candle aspect of AGB carbon stars and concluded
that the mean $I$-band magnitude of C-stars can provide reliable
distance determinations up to $\sim 2$ Mpc with currently available
ground-based telescopes.}

A systematic census of AGB C-stars in Local Group (LG) galaxies was
made over the last few years using the four-filter technique with two
optical broad-band and two narrow-band filters
\citep{albe+2000,battdeme2000,nowo+2001}.

A different approach \citep[e.g.][]{cionhabi2005,gull+2005,kang+2005}
is based on the combination of one optical filter with $J$ and $K_s$
near-IR broad-band filters.
\referee{
In both near-IR broad-band and in optical broad plus narrow-band
filter techniques, C and M-type stars can be easily separated in
color-color diagrams. However the selection criteria are somewhat
different, and both methods are likely to miss some C-stars, or to
select some that would appear as M-type stars when considering the
alternative method \citep[e.g][and later in this
work]{deme+2006}. More specifically the advantages of near-IR over
purely optical techniques are: }
(i) extinction in the near-IR is much lower than in the optical,
reducing the problems with differential reddening that can be present
in star forming dwarf irregular (dIrr) galaxies;
(ii) the spectral energy distribution (SED) of AGB stars peaks in the
near-IR and makes them easily distinguishable from red giant branch
(RGB) stars;
(iii) from $J - K_s$ color and $K_s$ magnitude it is possible to
derive precise bolometric magnitudes; and
(iv) with the development of adaptive optics at large telescopes (NACO
at VLT, adaptive optics at the next generation of Extremely Large
Telescopes), operating principally in the near-IR, it is important to
know well the observational characteristics and have well prepared
models for the interpretation of the data of more distant objects that
will shortly be observed. It is important to know how these bright
near-IR sources, the first to be observed in a distant galaxy, behave
with age and metallicity.

Following this approach we present the first $J$ and $K_s$ photometry
of SagDIG and DDO210, two southern Local Group metal-poor dIrr
galaxies.  Near-IR data are complemented with ground based $I$-band
photometry for DDO 210, and $V$, $I$ ACS/HST photometry by
\cite{moma+2005} for SagDIG.

\begin{table}
\caption{Main parameters of the observed galaxies. 
$^{(1)}$\cite{moma+2005};
$^{(2)}$\cite{lee+1999};
$^{(3)}$\cite{mcco+2006};
$^{(4)}$\cite{lo+1993}
}
\label{t:par}
\centering
\begin{tabular}{c r@{}l r@{}l}
\hline\hline
&\multicolumn{2}{c}{SagDIG}&
\multicolumn{2}{c}{DDO 210}\\
\hline
$(m-M)_0$   & $25.10$ & $^{(1)}$ &  $24.89$&$^{(2)}$\\
$E_{(B-V)}$ & $  0.12$ & $^{(1)}$  &  $  0.03$&$^{(2)}$\\
$\feh$& $-2.0$&$^{(1)}$ & 
$-1.9^{(2)}, \ -1.3$&$^{(3)}$\\
$M_v$&$-12.3$&&$-10.9$&$^{(2)}$\\
$M_{\rm HI}/10^6 M_\odot$&$8$&$^{(4)}$ &$3$&$^{(4)}$ \\
\hline
\end{tabular}
\end{table}

DDO 210, also known as Aquarius dIrr, is one of the faintest dwarf
galaxies \citep{lee+1999}, while SagDIG is one of the brightest (see
Table \ref{t:par}). The number of C-stars is roughly a function of the
galaxy mass or absolute magnitude
\referee{\citep{groe2004,battdeme2005}},
so the number of expected C-stars in the two galaxies is very
different.  As a part of a homogeneous survey, \cite{battdeme2000}
found only three C-stars in DDO 210, and \cite{demebatt2002} found 16 in
SagDIG.

The aim of this paper is to investigate the AGB stars of SagDIG and
DDO 210 and in particular to characterize the near-IR properties of
their C-star population.

\section{Observations}

\subsection{SagDIG}
Observations were made during August 2003 with the SOFI \referee{\citep{moor+1998}}
camera mounted on the NTT telescope at ESO/La Silla Observatory.  The
camera is equipped with a $1024 \times 1024$ Hawaii CCD.  In Large
Field mode, with a spatial resolution of $0 \farcs 288 \ {\rm
 pix}^{-1}$, the field of view is $ 4\farcm 9 \times 4\farcm 9$.  Our
data set consists of of 24 $J$ frames and 48 $K_s$ frames centered on
SagDIG (see Table \ref{t:log} for a log of the observations).  SagDIG is
close in projection to Galactic center, so the number 
of contaminating foreground stars expected in the color magnitude diagram 
(CMD) is high.  To
have an estimate of the contribution of the foreground stellar
population we also observed an external field, located beyond the
tidal radius of SagDIG.

$J$ and $K_s$ data from the ESO archive (prog
ID. 61.E-0273) were also used.
Observations were carried out with SOFI in Large
Field mode during two nights in August 1998 and  the data set consists of
18 $J$ frames and 108 $K_s$ frame centered on SagDIG.

\begin{table}
\caption{Observing Log.}
\label{t:log}
\centering
\begin{tabular}{c@{ }l  l r r r@{$\times$}l}
\hline\hline
\multicolumn{2}{c}{Night}&
Filter &
$N_{\rm TOT}$ &
$N_{\rm Used}$ &
DIT (sec) &NDIT \\
\hline
\multicolumn{7}{c}{SagDIG}\\ 
\multicolumn{7}{c}{
$\alpha\text{(J2000)}=19^h29^m59^s$,
$\delta\text{(J2000)}=-17\degr40\arcmin42\arcsec$}\\[.5em]
11 - 12    &Aug 1998	& $J$    &  18 &  9         &  20.0&6\\
11 - 12    &Aug 1998	& $K_s$   &  108&  108       &  10.0&6\\
1 - 2      &Aug 2003	& $J$    &  24 &  13        &  10.0&6\\
1 - 2      &Aug 2003	& $K_s$   &  48 &  31        &   6.0&10\\
\hline
\multicolumn{7}{c}{SagDIG control field}\\
\multicolumn{7}{c}{
$\alpha\text{(J2000)}=19^h29^m24^s$,
$\delta\text{(J2000)}=-17\degr32\arcmin22\arcsec$}\\[.5em]
1 - 2      &Aug 2003	& $J$    &  24 &  24        &  10.0&6\\
1 - 2      &Aug 2003	& $K_s$   &  48 &  48        &   6.0&10\\
\hline
\multicolumn{7}{c}{DDO 210}\\
\multicolumn{7}{c}{
$\alpha\text{(J2000)}=20^h47^m01^s$,
$\delta\text{(J2000)}=-12\degr51\arcmin00\arcsec$}\\[.5em]
1 - 2      &Aug 2003	& $J$    &  24 &  24        &  10.0&6\\
1 - 2      &Aug 2003	& $K_s$   &  48 &  48        &   6.0&10\\
1       &Aug 2003	& $I$   &  4 &   4     &   300.0&1\\
1       &Aug 2003	& $I$   &  2&     2   &   10.0&1\\
\hline
\end{tabular}
\end{table}

\subsection{DDO210}

Two contiguous fields in DDO 210 were observed with SOFI, 
following the same strategy adopted for SagDIG.
In addition, four deep and two shallow images in the $I$-band
were taken with EMMI, 
\referee{
the NTT multi-mode instrument
\citep{dekk+1986},
} in RILD mode.
The camera is equipped with two $2048 \times 4096$ detectors. 
In $2 \times 2$ binning mode the pixel scale is $0\farcs33 \ {\rm pix}^{-1}$ providing a 
field of view of $9\farcm9 \times 9\farcm 9$, so with one pointing we could cover both SOFI fields.
Observations are summarized in Table \ref{t:log}.

\section{Data reduction}
\subsection{SOFI data}
Following the SOFI User's manual we did not apply dark
frame correction, because the dark current is low and
{\it
``for broad-band imaging the dark frames is
a poor approximation of the underlying bias pattern, so subtracting it
serves no real purpose''}.

Special Flats frames were obtained multiplying the flat field
frames with the so-called {\it illumination correction} frames that were
downloaded from the SOFI webpages\footnote{
http://www.ls.eso.org/lasilla/sciops/ntt/sofi/images/fits/Archive/}.

Reduction of the data was done using standard IRAF scripts,
independently per night and filter,
following the recipes described by
\cite{rejk+2001}.

For each image, the sky subtraction was made using the
five frames closest in time as {\it ``sky''} frames and
creating object masks with SEXTRACTOR to 
avoid overestimation of the sky level.
The final sky frame was computed averaging all
five {\it ``sky''} frames with the objects masked.  The corresponding
final sky frame was subtracted from each image which was then divided by the
Special Flat frame.
Finally the mean value of the sky frame was added.

Given the relatively large number of SagDIG frames, only those with good
seeing (better than $1 \farcs 1$) were used. The resulting number of
frames that are used below is given in column 4 of
Table \ref{t:log}.  For DDO 210, we used all available frames.

Aperture photometry was done on each image using DAOPHOT
\citep{stet1987} and the relative shifts between all frames belonging to
a single epoch (1998 and 2003) were calculated with DAOMASTER \referee{\citep{stet1993}}. The
frames were finally registered with {\it imalign} and combined with 
{\it imcombine} tasks in IRAF.  The final result are the master $J$
and $K_s$ images of each epoch.

A posteriori we verified that rejecting the frames
with poorer seeing leads to deeper master images
with a higher S/N ratio.

\subsection{EMMI data}
The frames delivered by the EMMI CCD are in the Multi Extension Fits Format. To
read them and re-create a single image, a MIDAS script, provided by the La Silla
Observatory, was
used\footnote{http://www.ls.eso.org/lasilla/sciops/ntt/emmi/quickred/inemmi.prg}. 
Images were then processed within IRAF
using the {\it xccdred} package. The reduction steps included bias subtraction and
division by normalized twilight flat-field images.

\section{Photometry}
\subsection{SOFI data}
Point-spread-function (PSF) photometry was performed using
DAOPHOT/ALLFRAME \citep{stet1987,stet1994} on each single epoch master
image. The PSF was generated with a PENNY function with a quadratic
dependence on position on the frame.
The photometry of the two epochs
was finally combined with DAOMASTER.  We also produced a catalogue
from the external SagDIG field observations, following the same procedure
previously described.  In order to use the latter for
estimating the contribution of the foreground stellar population, we
generated a catalogue using only our observations (Aug 2003 run).

All master images were astrometrically calibrated using the
2MASS point source catalogue \citep{cutri+2003} as the reference.
The relative astrometric accuracy is always better than 
$0 \farcs 3$.

\realfigure{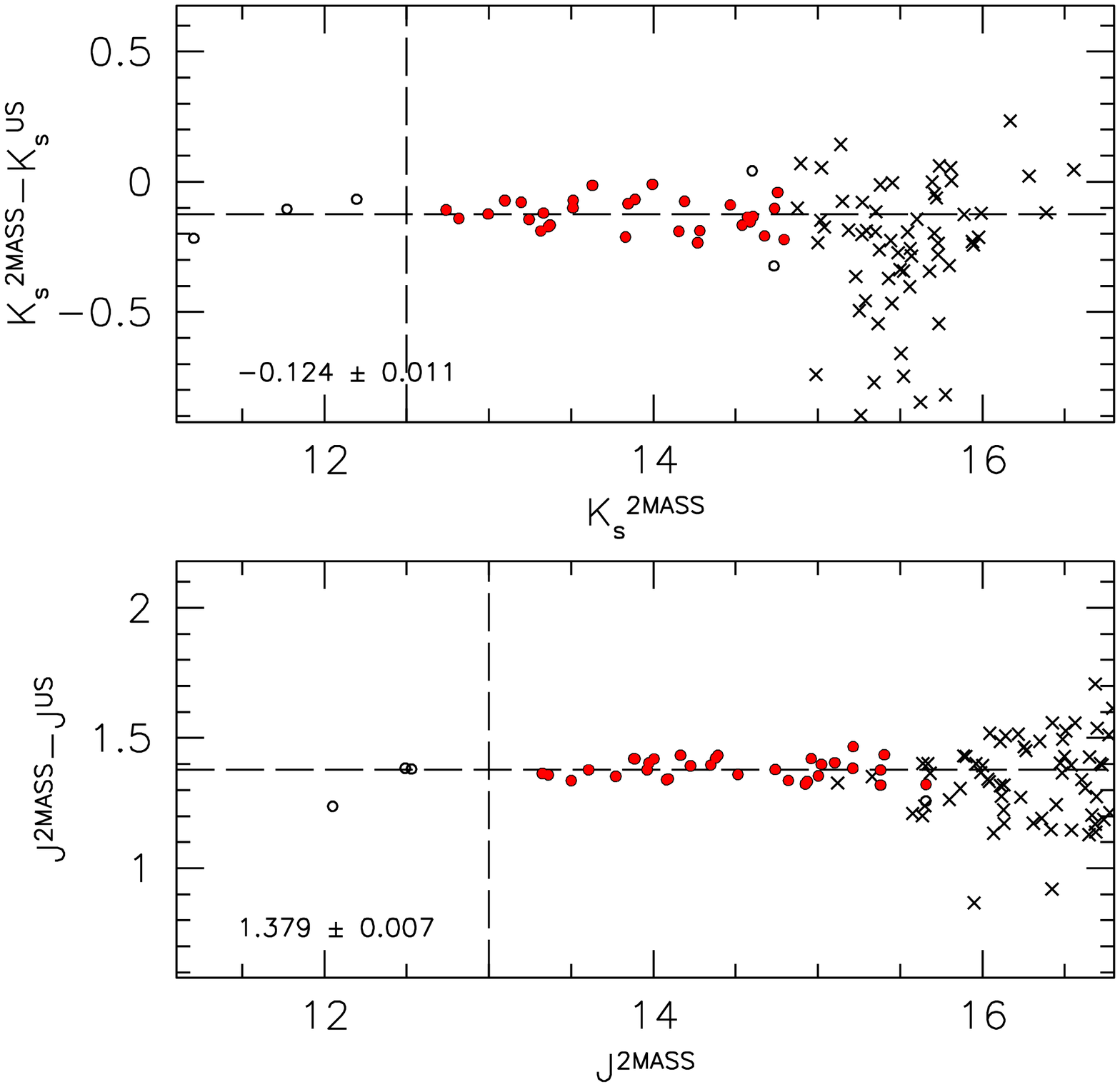}{
Zero point differences between our SagDIG uncalibrated magnitudes and the
2MASS point source catalogue. Crosses are 2MASS sources with $S/N<10$
that were not used. Stars marked as open circles were rejected after a $\kappa-\sigma$
clipping. We also rejected bright objects to avoid possible
saturation problems.}{f:calib}

The color term between the LCO photometric system defined by
\cite{pers+1998} and that of  SOFI is negligible \citep{gull+2006}. 
Moreover \cite{carp2001} shows that the
LCO and the 2MASS systems are equivalent.
Given these, photometric calibration was done applying the zero point
difference between our uncalibrated catalogue and the 2MASS point
source catalogue \citep{cutri+2003}, as shown in Fig. \ref{f:calib} for SagDIG.
Photometric calibration of DDO~210 SOFI data was done in the same way.

\subsection{EMMI data}
PSF fitting photometry was performed, using standard DAOPHOT/ALLFRAME
procedures. The PSF was generated with a PENNY function with a quadratic
dependence on position on the frame. The list of objects used as input for ALLFRAME was 
generated on a median image produced from all six EMMI frames. The final ALLFRAME run was
performed on all six frames, and the resulting catalogues were generated with DAOMASTER.
\referee{Instrumental magnitudes have been corrected for aperture corrections. These corrections have been derived for each EMMI chip separately using several bright and isolated stars.}
The photometry was calibrated using the solution derived by \cite{rejk+2005} for 
the same observing night:
\begin{equation}
i = I - 25.614+ 0.054 \times  {\rm Airmass} \quad [+ 0.040 * (V-I)]
\end{equation}
where $i$ are \referee{aperture-corrected instrumental magnitudes 
and $I$ are magnitudes calibrated to the standard Johnson/Kron-Cousins photometric system.}
We have no $V$-band observations, so the $(V-I)$ color term was ignored. The resulting calibration
of the $I$ photometry is indeed uncertain, but for our purpose this is not a major limitation. The optical
data of DD0 210 are used only marginally and our conclusions are not affected by the 
accuracy of the EMMI $I$-band photometric calibration.
\referee{However the error
in the $I$ magnitude due to the color  term should not exceed $0.1$ for a single
star. Considering also the uncertainties on the aperture correction and on the calibrating relation,
we estimated that the final error associated to $I$-band is $0.15$ mag.}

\section{Color-magnitude diagrams}
\subsection{DDO210}\label{s:cmdddo}

\begin{figure*}
\centering
\includegraphics[width=8.2cm]{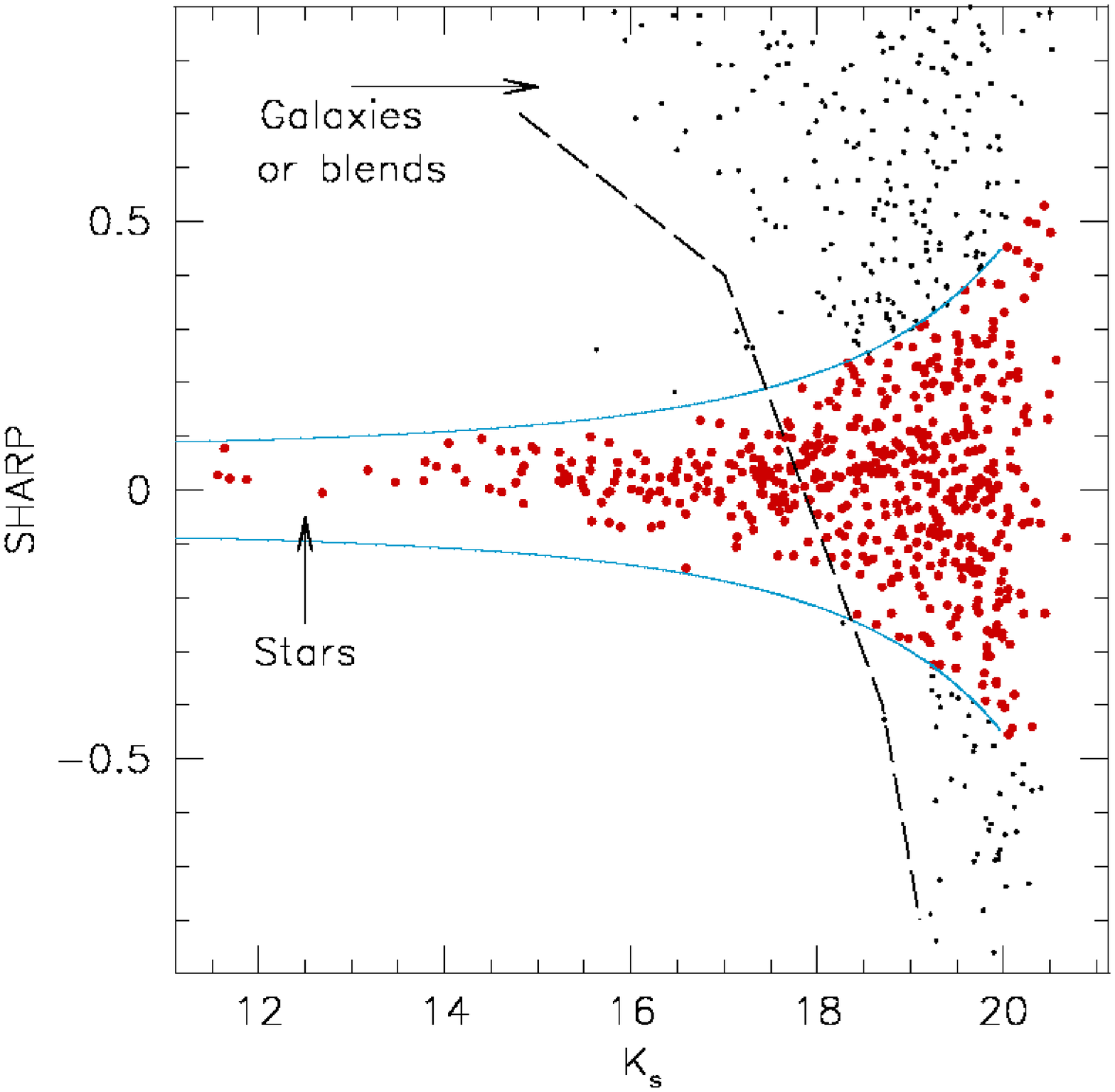}\hspace{4mm}
\includegraphics[width=8.2cm]{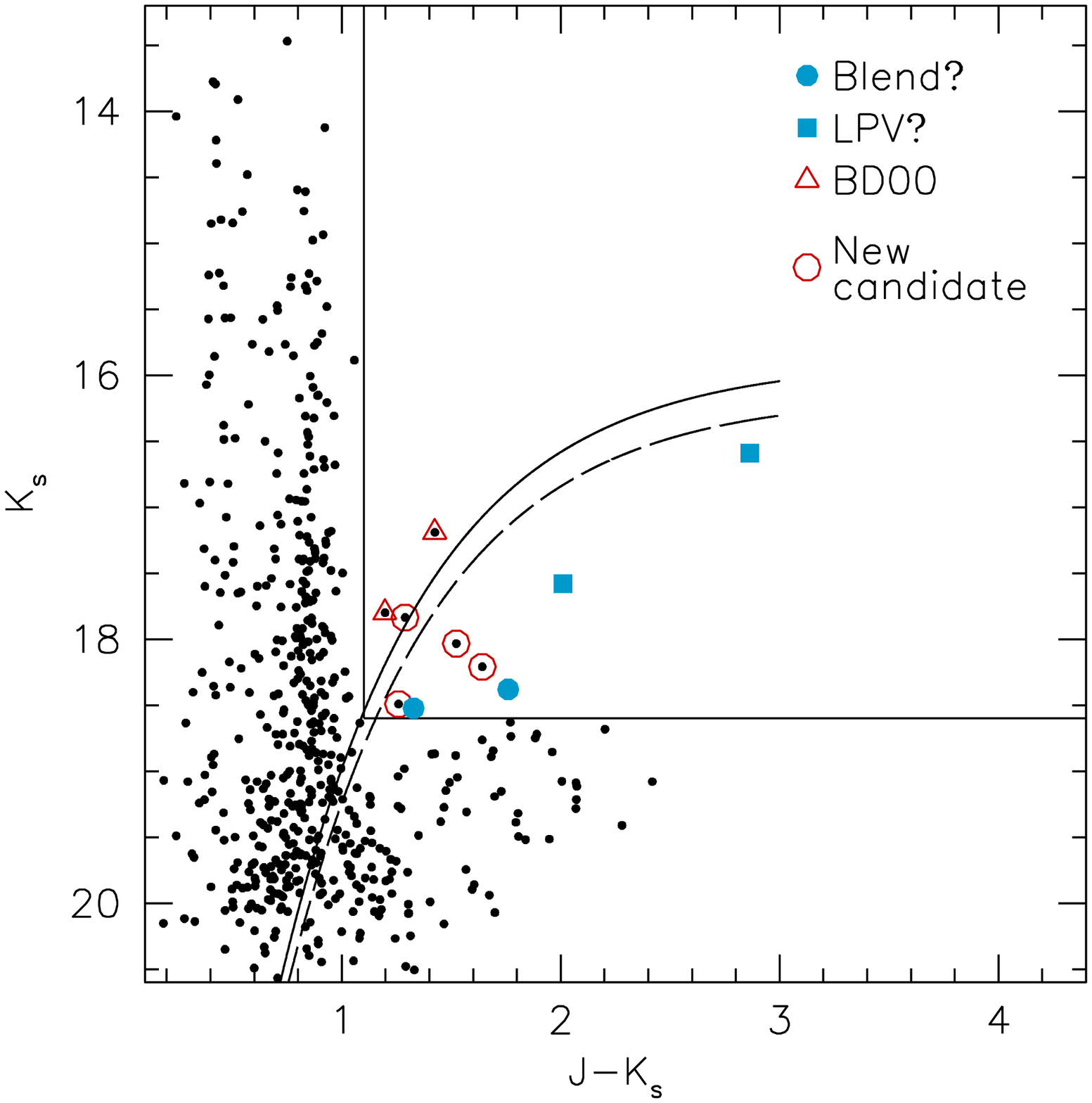}
\caption{
\referee{
{\it Left panel:}
The SHARP parameter as a function of $K_s$ magnitude. Solid lines
show the criterion adopted to select stellar objects ({\it larger
dots}).  Dashed lines indicate the limits of the region occupied by
unresolved galaxies and photometric blends.}
{\it Right panel:}
The near-IR CMD in the field of DDO 210 and the carbon stars
identified by \cite{battdeme2000}.  Probable photometric blends are
shown as {\it filled circles}, while \rereferee{{\it squares}} are candidate
LPVs. Solid lines show the selection criterion for candidate
C-stars.
\referee{The four new candidates are shown by {\it empty circles}.
For comparison we also show  the mean carbon star color-magnitude
relation from Totten et al. (2000) assuming a distance modulus of
24.89 \citep[{\it solid line}]{lee+1999}
25.15 \citep[{\it dashed line}]{mcco+2006}
}
}\label{f:ddodemers}
\end{figure*}

\referee{We have selected stellar objects using the SHARP image
  quality diagnostic provided by ALLFRAME.  Isolated bright stars have
  a shape very similar to the adopted PSF and a SHARP value close to
  zero. Unresolved galaxies with elliptical shapes or unrecognized
  blended stars have SHARP values significantly greater than zero.
  Cosmic rays and bad pixels have SHARP values below zero. The SHARP
  value for all objects in our catalogue are plotted in left panel of
  Fig. \ref{f:ddodemers} as a function of $K_s$ magnitude.  The
  increase of the spread of SHARP values of stellar objects at fainter
  magnitudes is due to the lower S/N ratio at these magnitudes. On the
  other hand, the size of fainter unresolved galaxies is smaller and
  some of these could have shapes and sizes similar to the stellar
  PSF.  As a consequence, the sequences of stellar objects and
  unresolved galaxies/blends merge together at $K_s\sim19$.  The solid
  line in Fig. \ref{f:ddodemers} shows the function we defined to
  select stellar objects. At fainter magnitudes, contamination by some
  unresolved galaxies/blends is possible, but a more severe selection
  would exclude some faint stars with low S/N ratio.}

The near-IR $JK_s$ CMD of DDO 210 is shown
in Fig. \ref{f:ddodemers}.  The two almost-vertical sequences at $J-K_s\sim 0.6$ and
$J-K_s\sim 0.9$ are populated by foreground Galactic stars
\citep[e.g.][]{nikowein2000}.  
Adopting the parameters reported in Table~\ref{t:par} and the calibration of 
\cite{vale+2004}  the tip of the RGB (TRGB) is expected to be found at 
$J-K_s=0.7$ and $K_s=19.2$
\footnote{\rereferee{
The calculations were made using the global metallicity \mh, which is 
related to the \feh\ iron abundance by taking into account the 
$\alpha$ elements overabundance [$\alpha$/Fe] \citep{sala+1993}:
\begin{equation}
\mh=\feh + \log \ (0.638 f_{\alpha} +0.362); \quad  
f_{\alpha}=10^{\rm{[}\alpha \rm{/Fe]}.}
\end{equation}
The choice of \mh\ is motivated by the fact that this is considered a 
better metallicity indicator than \feh\ when comparing stellar populations
with different [$\alpha$/Fe], like GGCs and dwarf galaxies.
}}.
\referee{The latter was calculated assuming
\mh$=$\feh\ ([$\alpha$/Fe]$=0$)
and the \cite{rieklebo1985} reddening law.}
The photometric errors and the contamination from foreground stars are very high 
in the region of the RGB, so it is not visible in the CMD.

\referee{In near-IR CMDs oxygen-rich M giants are found 
along a sequence that is aligned with the RGB brighter than the RGB tip.
Carbon stars are found along a red tail which is clearly separated from M-stars and foreground stars sequences \citep[e.g.][]{cionhabi2005,kang+2005,gull+2006}. 
We can therefore conclude that stars found at colors significantly redder than 
the sequence at $J-K_s\sim 0.9$ in our CMD, are candidate C-stars.
The only other objects that could be found in this region are red background galaxies, but
these should be found at relatively faint magnitudes.
Some AGB stars are surrounded by a dusty circumstellar envelope lost by stellar winds and therefore
are found at redder color than normal AGB stars. Indeed, some red stars in our CMD could be
dust-enshrouded  M-stars.
Without additional low resolution spectroscopy they are indistinguishable from C-giants. 
However, in low metallicity populations such as those present in DDO210 and also in SagDIG, 
one expects to find a much higher frequency of C-rich with respect to O-rich AGB stars,
\citep[e.g.][]{groe2004}.
}

\cite{battdeme2000} found three carbon stars in DDO 210. One of them is outside our field of view;
the other two are reported in Fig. \ref{f:ddodemers}. 
\referee{
For comparison, the main locus of C-stars in LG dwarf galaxies obtained by
\cite{tott+2000} is also shown. We scaled it to the distance of DDO~210, adopting both the 
distance derived by \cite{lee+1999}
and the higher value derived by  \cite{mcco+2006}.
In our CMD there are stars other than the two C-stars listed by \cite{battdeme2000}
in the region where C-stars are expected to be found.
We assumed as C-stars candidates all objects in our CMD with
$J-K_s>1.1$ and $K_s<18.6$.
As shown in Fig. \ref{f:ddodemers}  there are 
ten objects within this region, the two \cite{battdeme2000} C-stars and eight objects that } 
can be considered good carbon stars candidates.
Nonetheless, to confirm them, higher spatial
resolution data, possibly taken with the HST/ACS, are required 
to distinguish them from background galaxies or photometric blends.
The more detailed analysis of SagDIG red stars presented in the following sections
shows that SOFI near-IR imaging is a very powerful tool for detecting C-stars, but 
completing the analysis with ACS optical imaging greatly improves the quality
and reliability of the conclusions.

\begin{figure*}
\centering
\includegraphics[width=8.2cm]{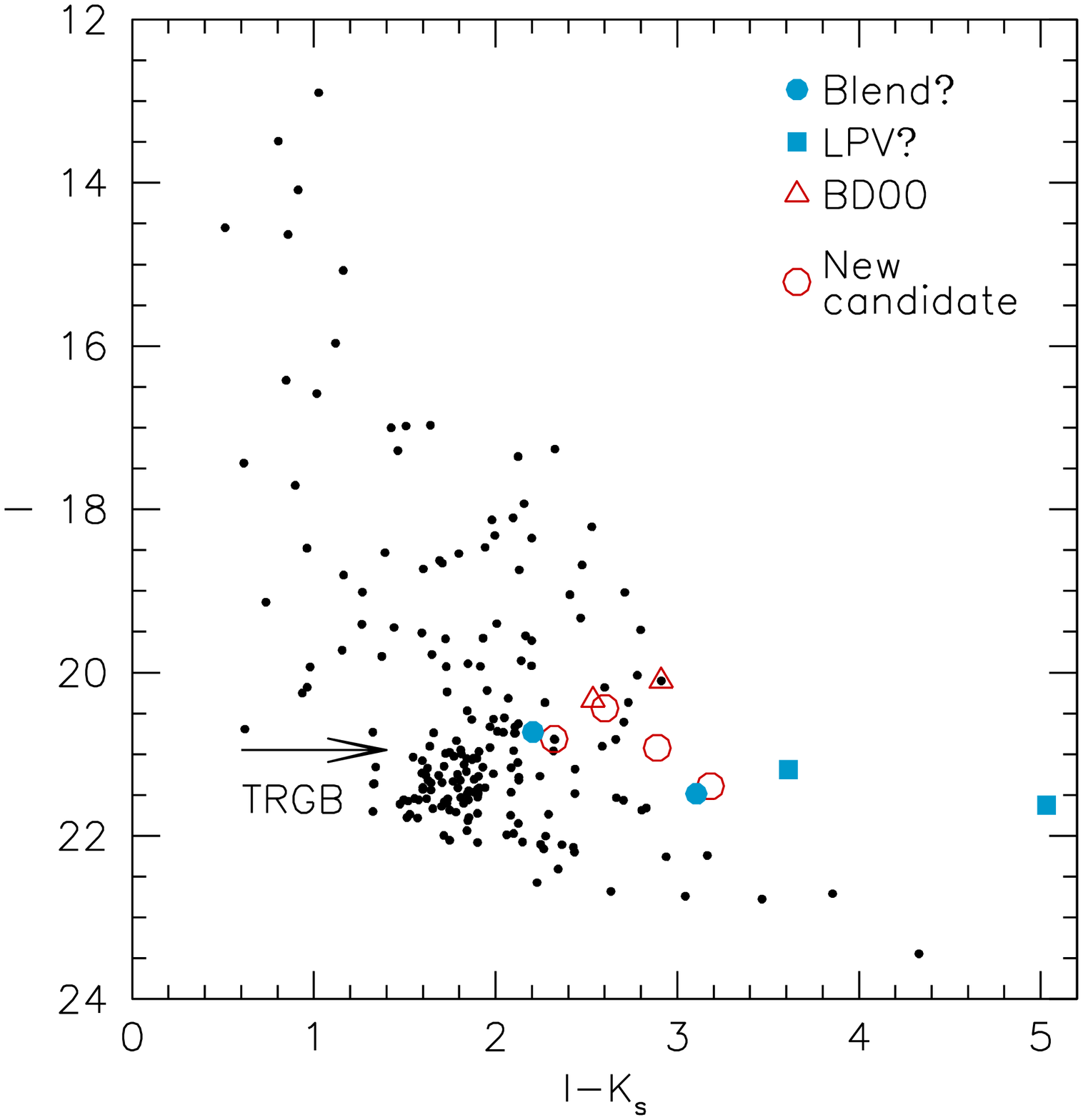}\hspace{4mm}
\includegraphics[width=8.2cm]{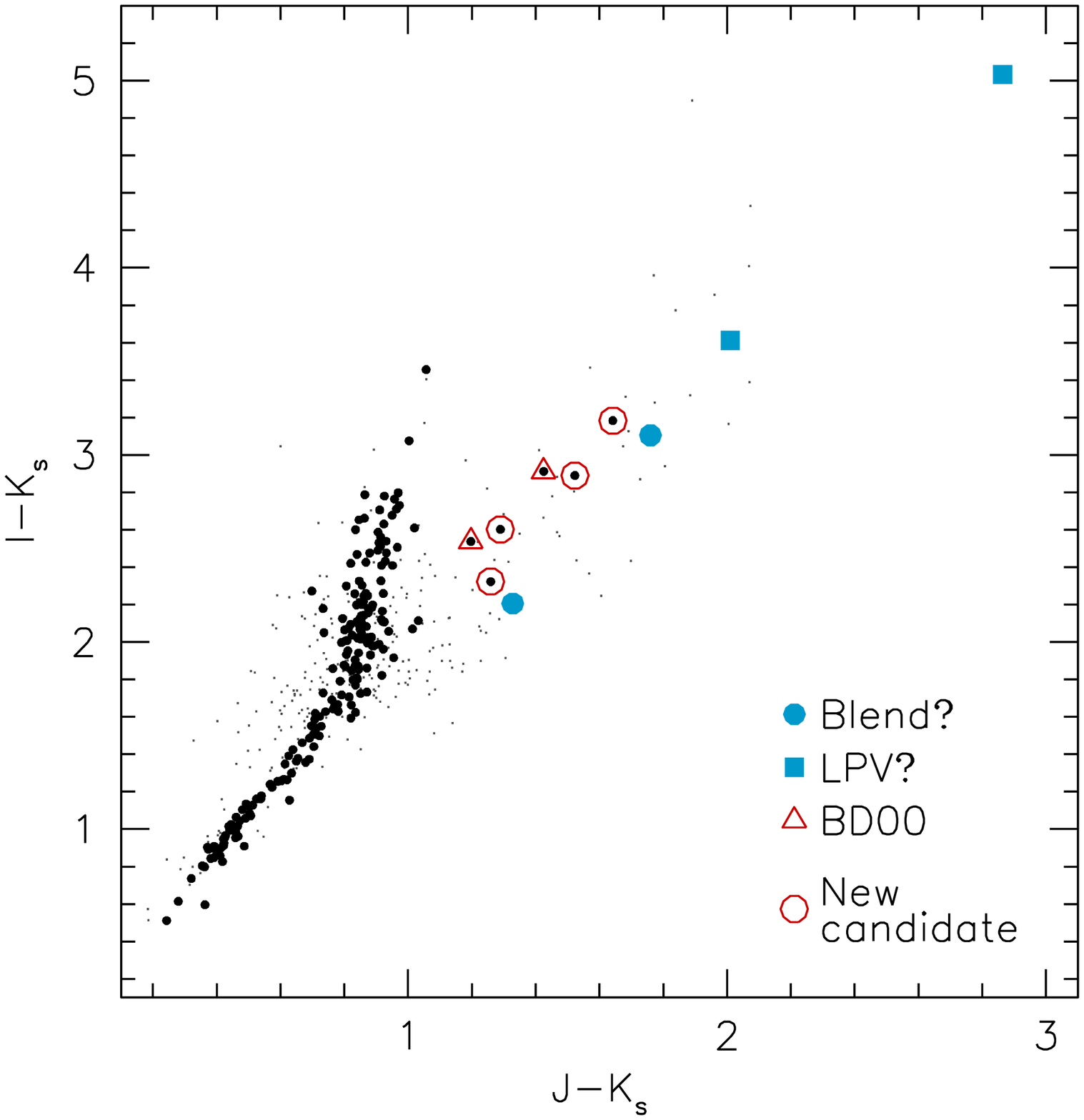}
\caption{
{\it Left panel:} optical - near-IR CMD of DDO 210 stars within $2\arcmin$ from the center. 
\referee{The outer regions are omitted from this plot for clarity.}
The arrow shows the position of the TRGB as determined by \cite{lee+1999}.
One of the two C-stars identified by \cite{battdeme2000}, shown as triangles, 
is located at a greater distance, and is not present in this CMD. 
\referee{The symbols are the same as in Fig. \ref{f:ddodemers}.}
{\it Right panel:} The two-color diagram of DDO210. Larger (darker) dots are 
stars with $K_s<18.6$.}
\label{f:ddott}
\end{figure*}

In Fig. \ref{f:ddott}  the near-IR data are combined with the $I$-band photometry.
In the left panel the upper  RGB is visible. \cite{lee+1999} found the TRGB at 
$I_{\rm TRGB}=20.95\pm0.10$. The calibration of our $I$ magnitudes has larger uncertainty, 
but nevertheless
the TRGB position in our CMD is in good agreement with the \cite{lee+1999} measurement.
In the right panel, the two-color diagram is presented. The almost vertical sequence of stars 
with $J-K_s\simeq0.9$ is populated by foreground Milky Way stars. The red sequence
at $J-K_s>1.1$ and  $I-K_s>2.2$ is the region where the carbon stars are expected 
\referee{\citep[e.g.][]{cion+2004}}
and in fact the two carbon stars identified by \cite{battdeme2000} are located in this region.
We point out that all the other eight red objects are found along this sequence. In particular there is one very red object 
with $J-K_s=2.9$, which, if confirmed, could be an AGB star enshrouded in an extended envelope of dust ejected by stellar winds.
The coordinates and photometry of the candidates carbon stars are listed in Table \ref{t:ddored}.

\begin{table*}
\caption{DDO 210 stars with $J-K_s>1.1$ and brighter than $K_s=18.6$.
In the last column the cross identification with \cite[BD00]{battdeme2000} carbon stars 
list is reported.
Stars marked with {\it LPV} are candidate long period variables, 
while stars marked with {\it Blue} are too blue in optical bands and are likely
photometric blends (see text for more details). 
For candidate LPVs the photometric variation is reported.
}
\label{t:ddored} 
\centering 
\begin{tabular}{c r@{$^h$} r@{$^m$} l@{\fs} l  r@{\degr} r@{\arcmin} l@{\farcs} l c c c c}
\hline \hline
ID &
\multicolumn{4}{c}{$\delta (2000)$}&
\multicolumn{4}{c}{$\alpha (2000)$}&
$K_s$&
$J-K_s$&
$I-K_s$&
note\\
\hline 
 1  & 20 & 46 & 47&79 & $-$12 & 50 & 51&77 &   18.38 &    1.76 &    3.10 &Blue\\
 2  & 20 & 46 & 46&51 & $-$12 & 52 & 57&03 &   18.52 &    1.33 &    2.21 &Blue\\
 3  & 20 & 46 & 39&22 & $-$12 & 50 & 31&36 &   17.80 &    1.20 &    2.54 &BD00 C02\\
 4  & 20 & 47 & 06&34 & $-$12 & 50 & 59&51 &   17.84 &    1.29 &    2.60 &\\
 5  & 20 & 47 & 04&23 & $-$12 & 49 & 54&10 &   18.21 &    1.64 &    3.18 &\\
 6  & 20 & 47 & 01&60 & $-$12 & 52 & 51&76 &   18.03 &    1.52 &    2.89 &\\
 7  & 20 & 46 & 53&92 & $-$12 & 50 & 37&68 &   16.59 &    2.86 &    5.03 & LPV? $\Delta I=-1.59$\\
 8  & 20 & 46 & 53&24 & $-$12 & 50 & 25&31 &   17.58 &    2.01 &    3.61 & LPV? $\Delta I=0.45$\\
 9  & 20 & 46 & 52&87 & $-$12 & 50 & 25&62 &   17.19 &    1.43 &    2.91 &BD00 C03\\
10 & 20 & 46 & 52&50 & $-$12 & 50 & 10&68 &   18.49 &    1.26 &    2.32 &\\
\hline 
\end{tabular}
\end{table*}

During the preparation of this manuscript \cite{mcco+2006} published an optical study of the stellar content of DDO~210. From the mean $I$-band magnitude of the stars located in a red clump they derive an average age of $4^{+2}_{-1}$~Gyr for the majority of stars in DDO~210. 
\referee{They also derived a mean age-corrected metallicity of  \feh\ $=-1.3$, which is much higher 
than the value \feh\ $=-1.9$ derived by \cite{lee+1999} and reported in Table \ref{t:par}.
These results suggest} that a significant population of carbon-rich AGB giants is expected to be present in this galaxy. In their CMDs they identify a bright asymptotic giant branch population. This further confirms the reality of our carbon-star candidates.

It is acknowledged that most, if not all, of the carbon giants vary on long time-scales 
\citep[e.g.][]{groe2004,raim+2005}.
 We therefore tried to search for variable star candidates by comparing the $I$-band photometry presented in this work with that of \cite{mcco+2006}. With only two epochs (separated by 734 days) it is possible to give only a lower limit to the number of long period variable (LPV) stars. It is however encouraging that two out of eight new carbon-star candidates are also candidate LPV stars.
\referee{The differences in magnitude between the two epochs are reported in Table \ref{t:ddored}. 
We note that these are lower limits to the variability amplitudes and that they are significantly
larger than the combined photometric errors at the corresponding magnitude.}
Obviously to confirm the variable nature of these stars, and to be
able to reliably estimate  the number of LPVs in this galaxy, more
observations are necessary. The two carbon-stars previously identified
by \cite{battdeme2000} do not present variability between the two
epochs to within the errors of photometric measurements.
\referee{Finally we note that the amplitude of LPV light curves increases at increasing colors 
\citep[e.g.][]{raim+2005} and in fact  
our two candidate LPVs are the two reddest candidate C-stars.
}
 
Finally we checked the position of all ten carbon star candidates in
our $I-K_s$ vs.\ $I$ CMD (Fig. \ref{f:ddott}) and in optical CMDs
using \cite{mcco+2006} data.  Eight of them, including the two
candidate LPVs and the two stars classified as C-stars by
\cite{battdeme2000}, are in the expected location for AGB stars.  The
other two, marked in Fig. \ref{f:ddodemers} (filled circles) and Table
\ref{t:ddored} (``Blue''), are too blue in optical bands
\rereferee{(e.g. in $B-V$)}.  They are similar to the three blue
objects found among SagDIG C-stars candidates (see next sections), with
blue colors in the optical and red colors in the near-IR, and thus it
is possible that they could be blends of two or more unresolved stars
or background galaxies.

\referee{\cite{battdeme2000} claim that their C-star candidate in
  DDO~210 may be unusually bright.  In Fig. \ref{f:ddodemers} our CMD
  is compared with the mean color-magnitude relation for C-stars
  derived by \cite{tott+2000} as a fit to the photometry of a sample
  of carbon stars from Milky Way satellite galaxies.  We assumed the
  distance modulus $(m-M)_0=20.89$ obtained by \citep{lee+1999}, but
  we used also the longer distance $(m-M)_0=25.15$ derived by
  \cite{mcco+2006}.  In both cases the C-stars detected by
  \cite{battdeme2000} appear to be slightly brighter than the
  \cite{tott+2000} relation. Interestingly our new candidates bring
  the C-star population of DDO~210 in agreement with the
  \cite{tott+2000} locus.  }

Summarizing, the number of bona-fide carbon stars in DDO is nine. This
includes three stars previously known from \cite{battdeme2000}, and six new
candidates discovered in this work that have both optical and near-IR
colors consistent with carbon giants. Two additional red objects are
probable blends \referee{or unresolved background galaxies}.

\subsection{SagDIG}
\realfigure{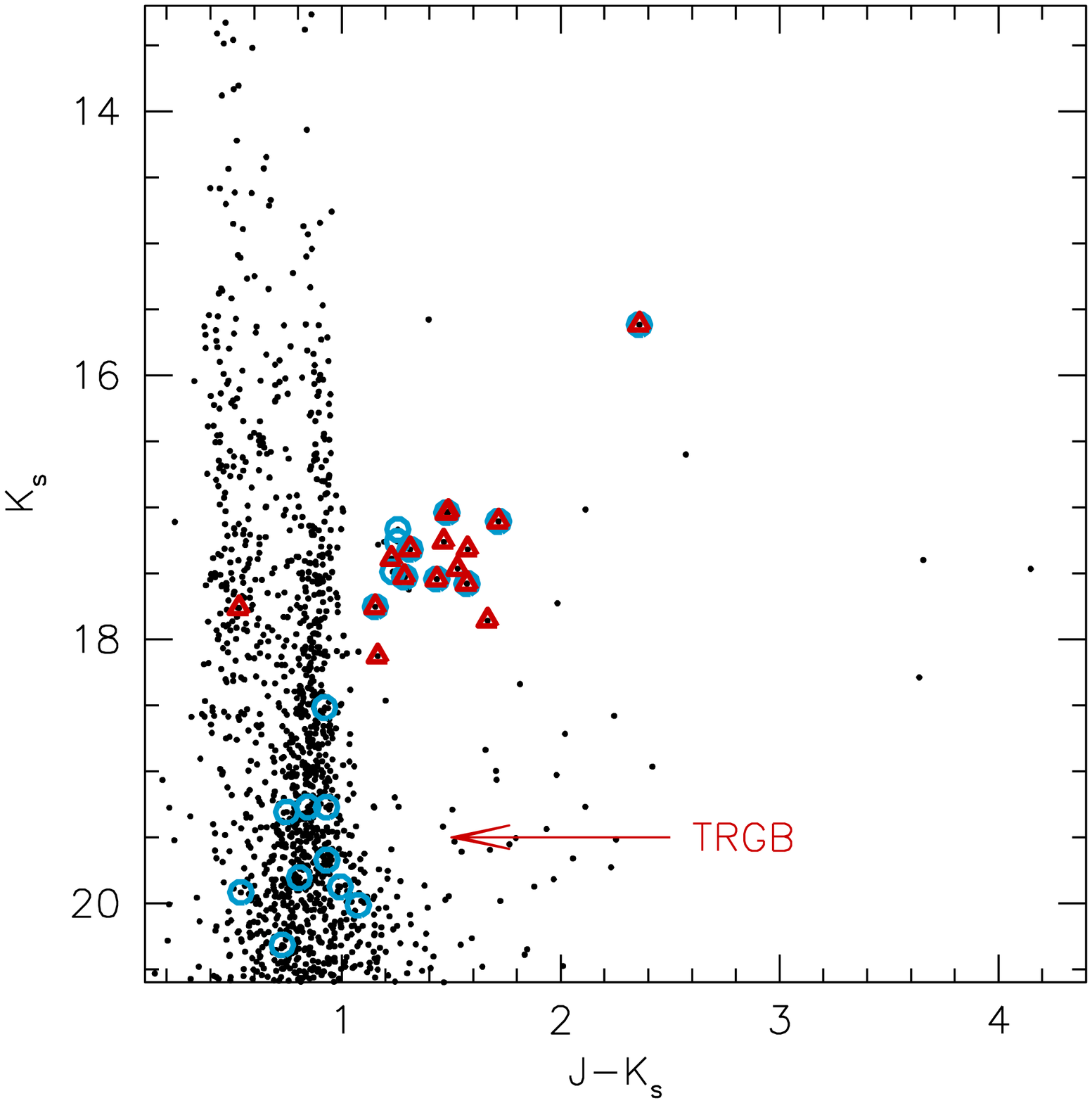}{
The near-IR CMD of SagDIG and the carbon stars identified by \cite{cook1987} (circles) and
\cite{demebatt2002} (triangles). Some of the carbon stars from \cite{cook1987} are fainter
than the TRGB, shown by the arrow.}{f:cmdsag}

The CMD of SagDIG is shown in Fig. \ref{f:cmdsag}.  
\referee{Stellar objects were selected using the SHARP parameter provided by ALLFRAME,
as described in Sect. \ref{s:cmdddo}.}
Also in this case the two almost vertical sequences at $J-K_s\sim 0.6$ and
$J-K_s\sim 0.9$, populated by foreground Galactic stars, are clearly visible
\citep[e.g.][]{nikowein2000}.  
\realfigure{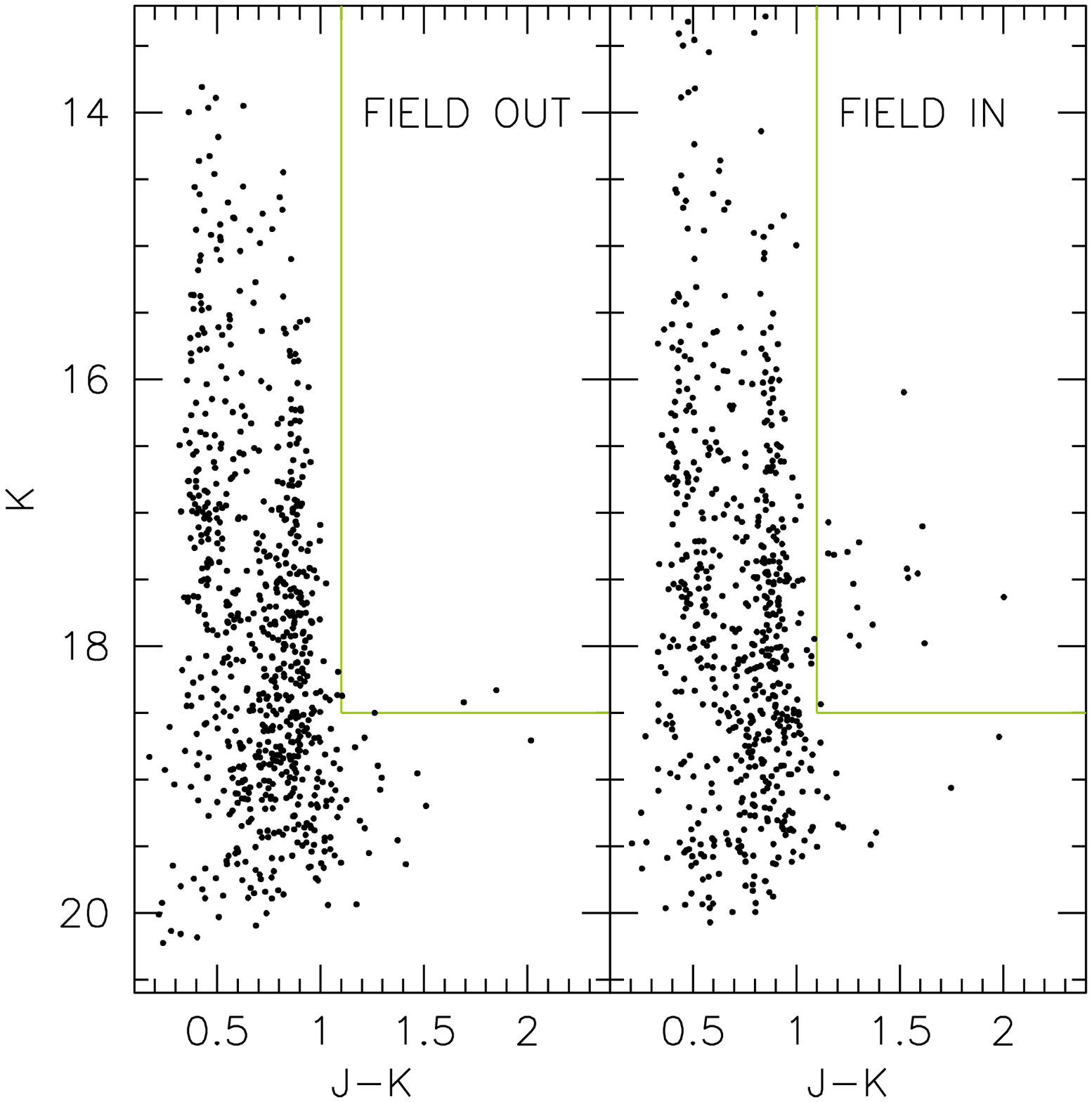}{
CMDs of the outer (left) and inner (right) SagDIG fields from August 
2003 observations. The box represents the region occupied by C-stars.}{f:fieldout}

In order to estimate the contribution of foreground Galactic stars we
compared the CMDs shown in Fig. \ref{f:fieldout}, obtained from the outer and the inner 
fields, but only from our observations (August 2003). In this way we compare data obtained
in exactly the same way, and observed under similar seeing conditions, 
so that the completeness and photometric errors in the two CMDs are similar.
\referee{The CMDs in Fig. \ref{f:fieldout} are obtained with the same SHARP selection defined to 
plot Fig. \ref{f:cmdsag}, which is obtained combining 1998 and 2003 observations.
The quality of photometry in Fig. \ref{f:fieldout} is clearly lower, and 
in particular, for the same star, the SHARP value is higher in the photometry obtained 
from 2003 observations alone.
For this reason some stars which are present in Fig. \ref{f:cmdsag} are absent in Fig. \ref{f:fieldout}.
}
The box in Fig. 
\ref{f:fieldout} corresponds to the region occupied by C-stars 
\referee{which is defined and discussed in detail below.}
No object is seen 
inside the box in the outer field
\referee{apart from two objects at the faint limit of the box, which are
likely background galaxies}.
Moreover a statistical analysis showed that the star counts in both
CMDs (excluding the C-stars region) are identical.
Adopting a distance modulus of 25.10 and 
\referee{\mh$=$\feh$=-2.0$},
using the relation of \cite{vale+2004} 
\referee{and the \cite{rieklebo1985} reddening law},
we derived that the expected magnitude for the tip of the red giant
branch is $K_s=19.5$, which is close to the limiting
magnitude in Fig. \ref{f:fieldout}.  We therefore conclude that all the stars
brighter than $K_s=19.5$ and bluer than $J-K_s=1.1$ are most probably
foreground stars and that all the stars redder than this limit are
candidate C-stars belonging to SagDIG.  The only possible exception
is the presence of AGB oxygen rich M-stars of SagDIG, which are expected to have
magnitudes brighter than the RGB tip, but colors similar to that of  foreground
stars \citep{nikowein2000}. However, given the low metallicity of SagDIG we 
expect a high C/M ratio and therefore a very small number of M-stars. This is confirmed
by the similarity of the luminosity function (LF) of the inner and outer fields in the color
range occupied by M-stars.

The CMD in Fig. \ref{f:cmdsag} obtained combining all the SagDIG data
(August 1998 + August 2003) is clearly deeper and extends to $K_s=21$.
The contamination from foreground stars is however high, and it is
impossible to distinguish the RGB of SagDIG.

We combined our data with the HST/ACS photometry of \cite{moma+2005},
transforming the ACS magnitudes
from the HST system to that defined by \cite{stet2000} (which is equivalent to the Landoldt standard system) using the equation given by
\cite{rejk+2005}:
\begin{align}\label{e:trasfACS}
V-F606W & = 0.222 \ (V-I) + 0.072\\ 
I-F814W & = -0.042 \ (V-I) + 0.130
\end{align}

\begin{figure*} 
\centering
\includegraphics[width=8.2cm]{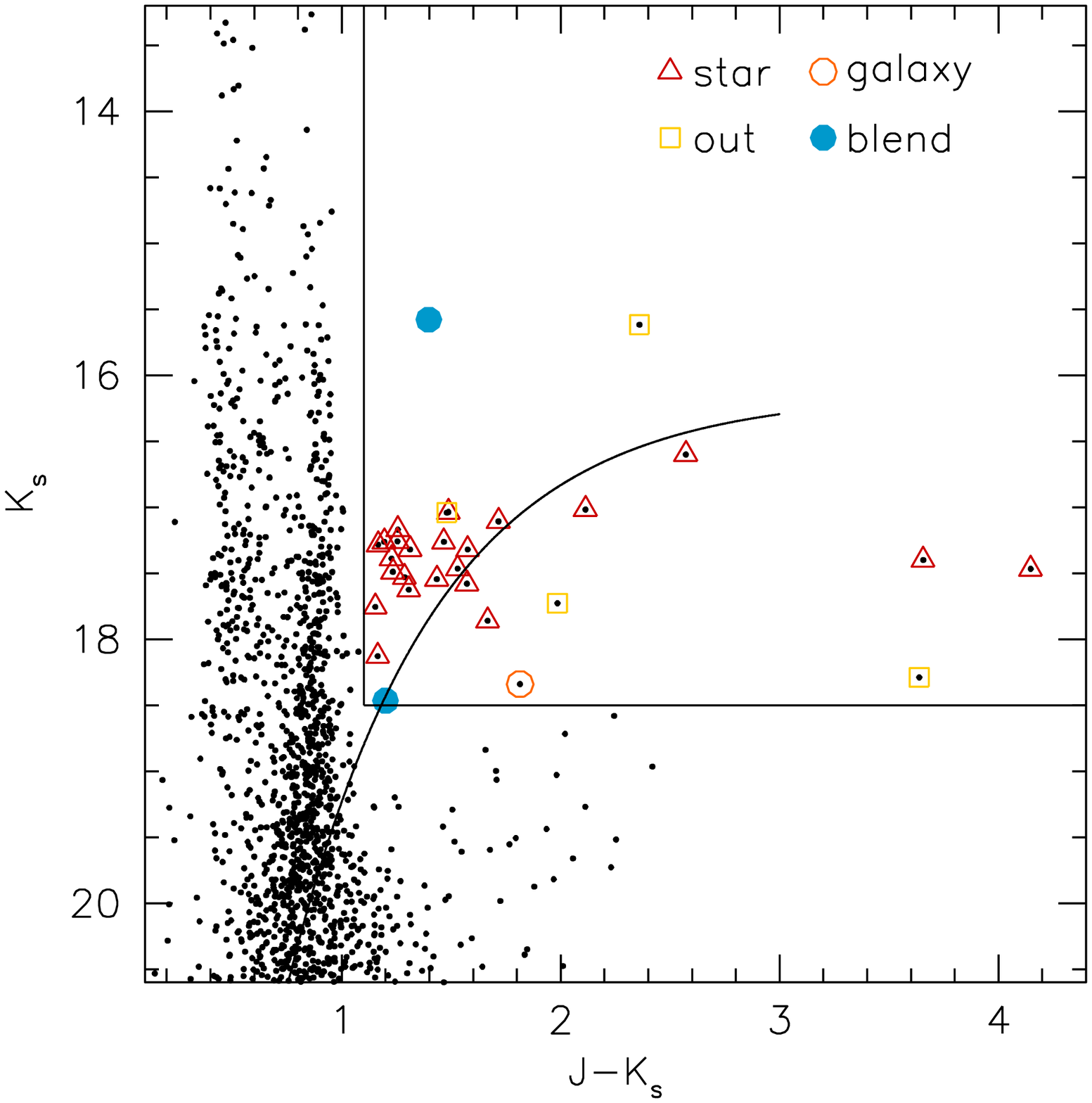}\hspace{4mm}
\includegraphics[width=8.2cm]{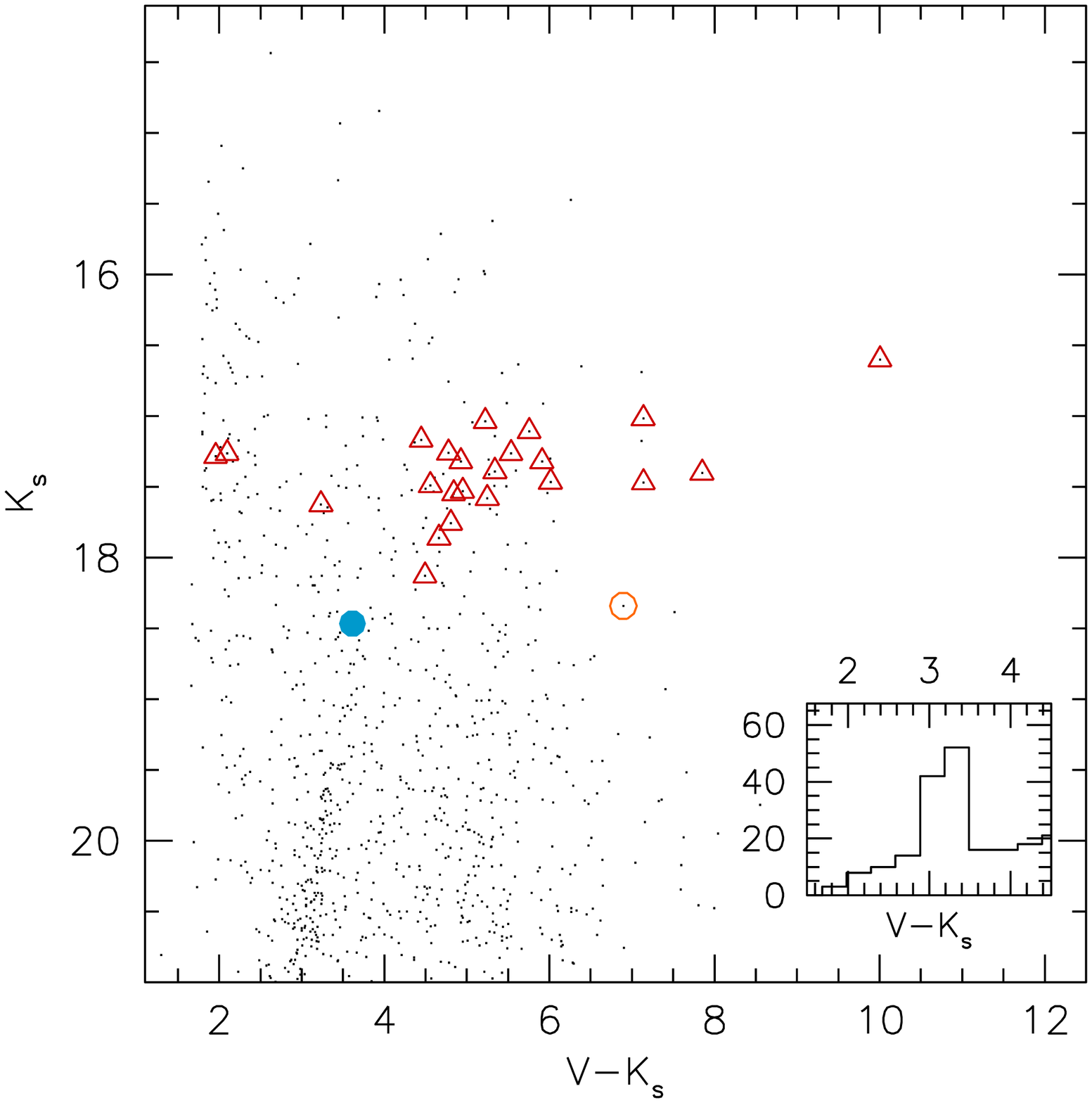}\\ \vspace{4mm}
\begin{tabular}{l@{\hspace{5mm}}r}
a)&b)\\
\includegraphics[width=8.2 cm]{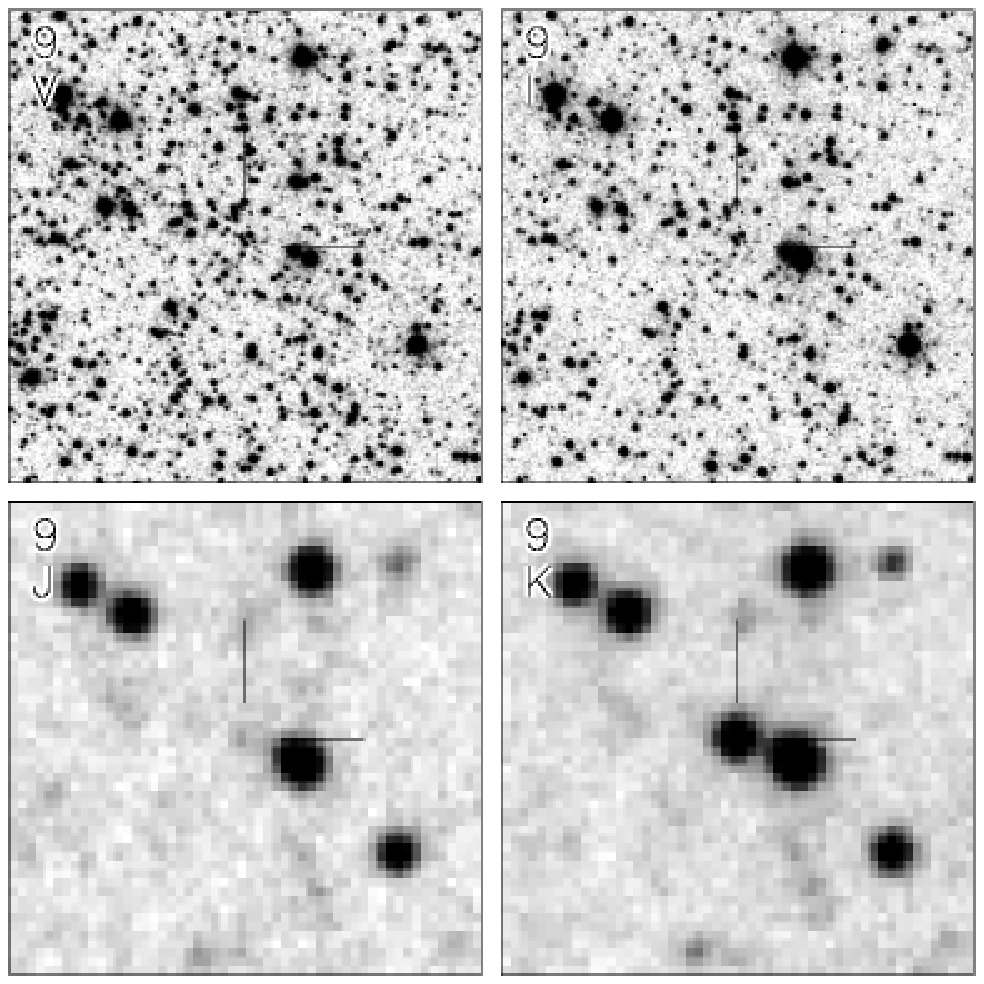}&
\includegraphics[width=8.2 cm]{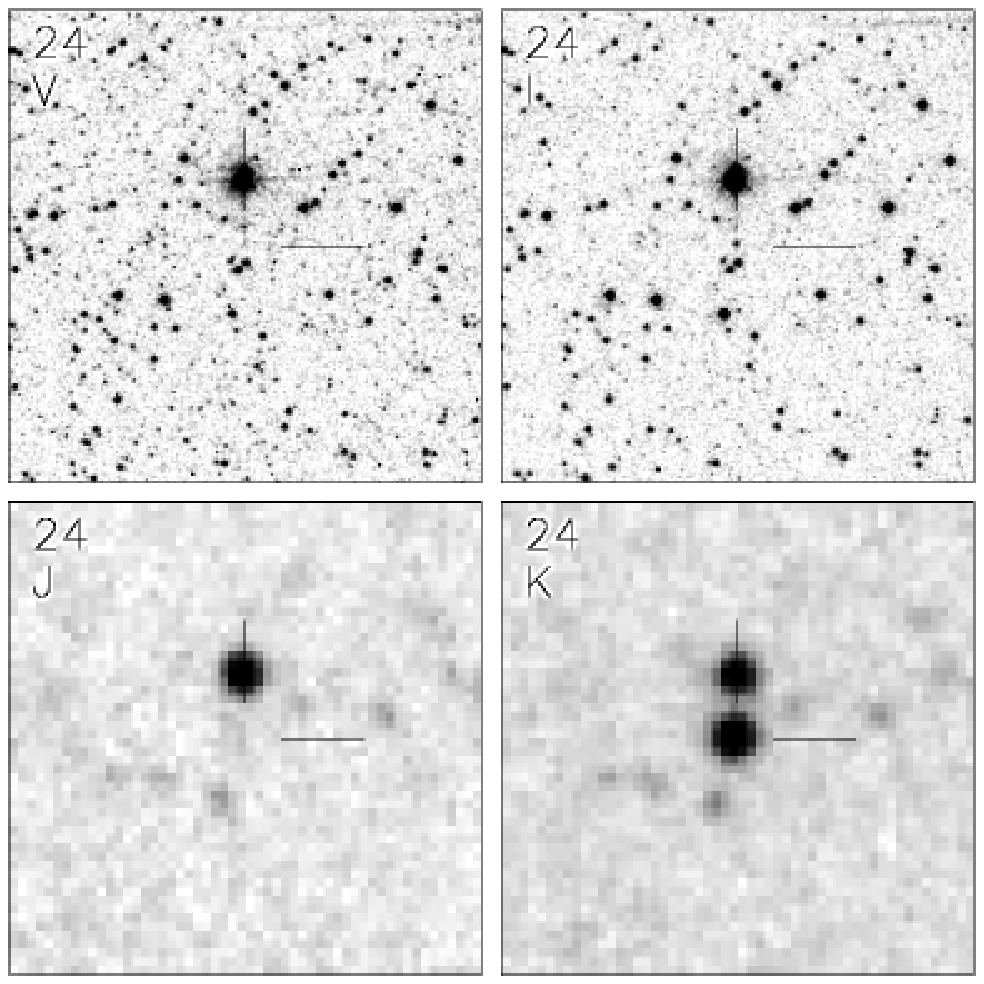}\\
c)&d)\\
\includegraphics[width=8.2 cm]{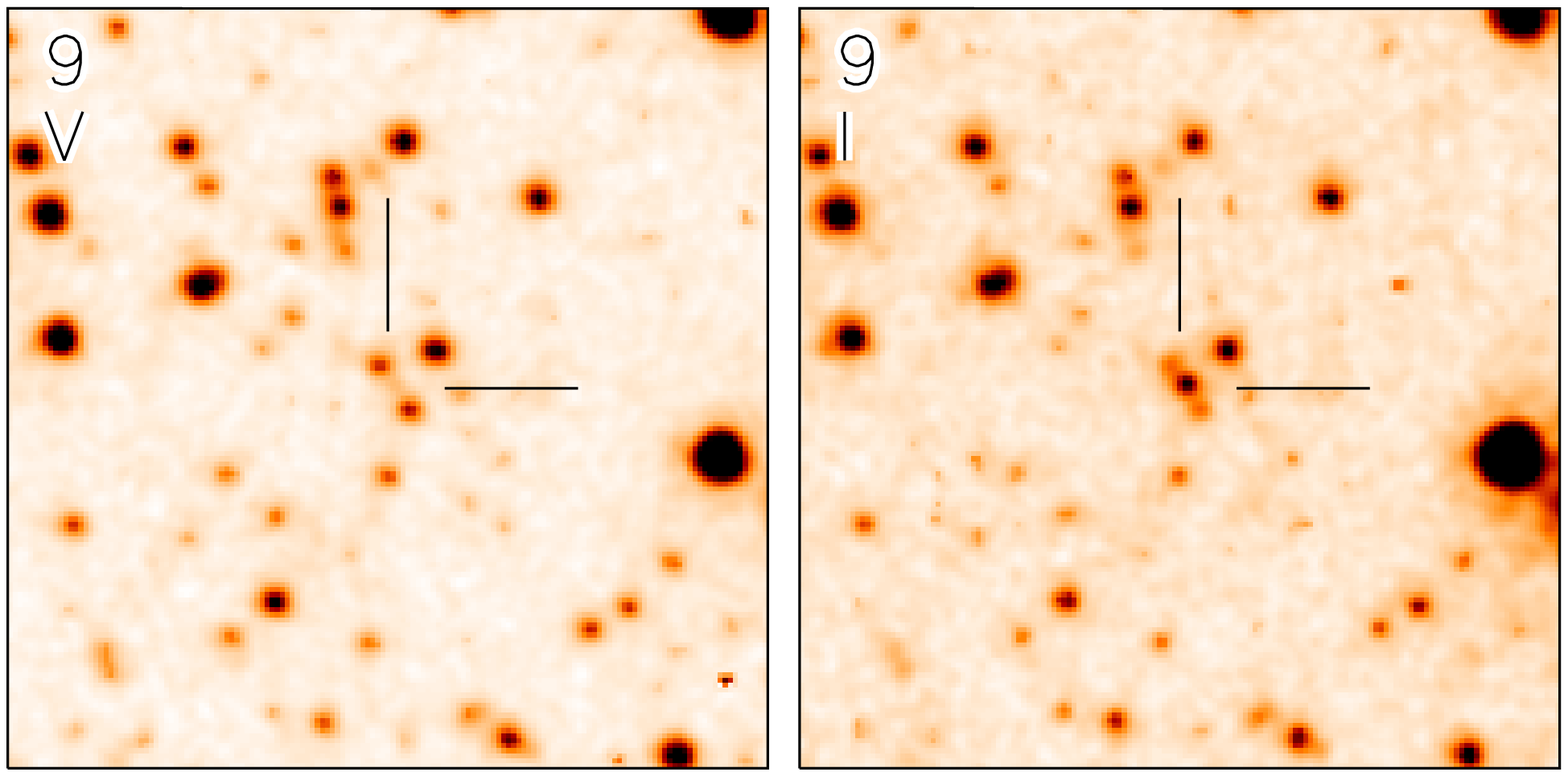}&
\includegraphics[width=8.2 cm]{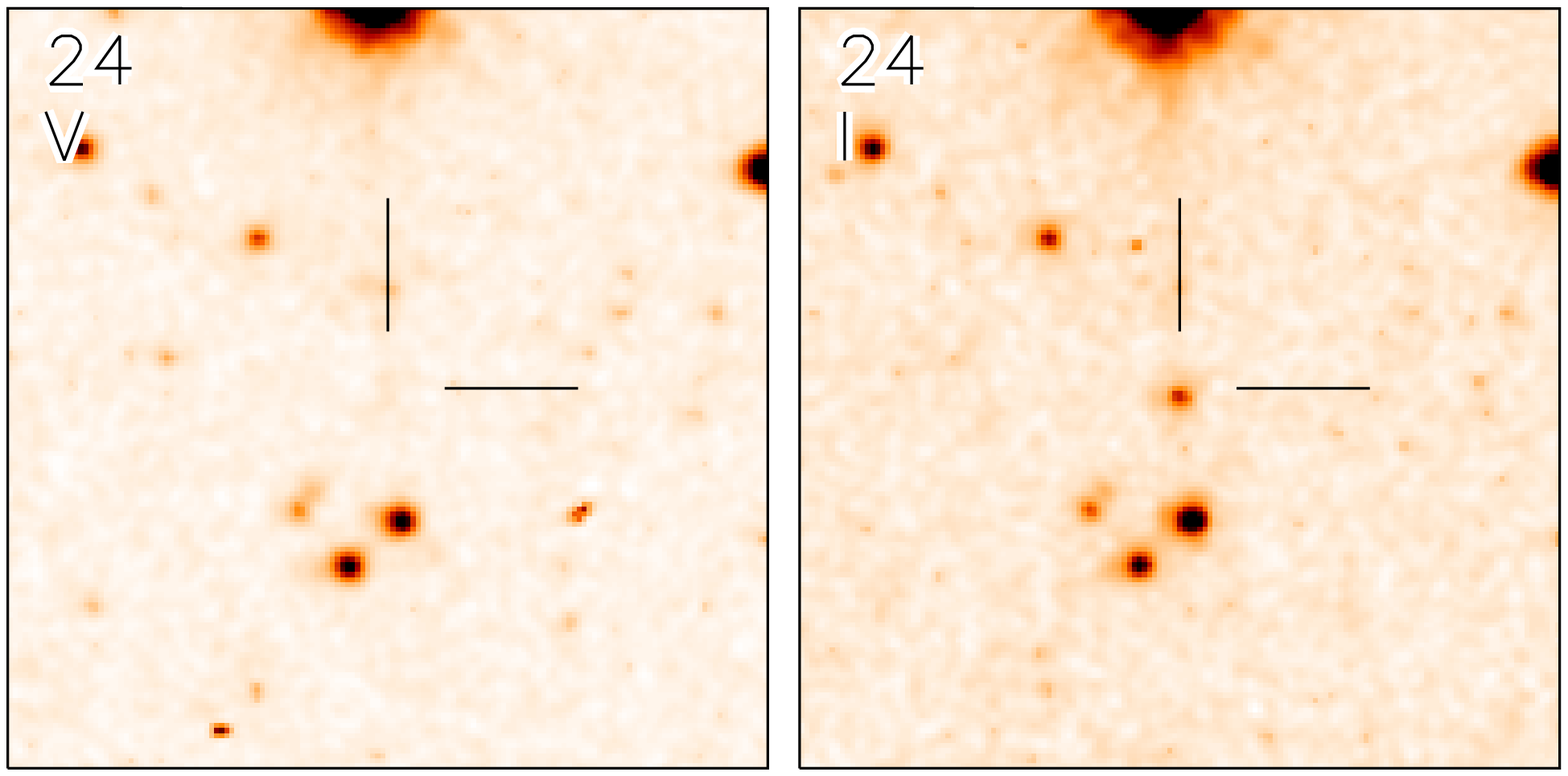}
\end{tabular}
%
\caption{
{\it Upper panels:}Classification of the red stars in our catalogue,
based on HST/ACS images \citep{moma+2005}. We plotted stars 
outside the field of view of the ACS observations as open squares.
\referee{
In the \rereferee{$K_s$}, $J-K_s$ \rereferee{CMD} we show the mean carbon star color-magnitude relation from Totten et al. (2000)
and the box we used to define C-star candidates.
The inset in the right panel shows the detection of the RGB in the $V-K_s$ color distribution
of stars with $20.5<K_s<19.5$.}
{\it Lower panels,}
{\it a) and b)}: Zoom in the regions of the two reddest stars (number 9 and 24) of SagDIG.  
Upper panels are HST/ACS $V$ and $I$ images and
lower panels are $J$ and $K_s$ SOFI images. The images show areas of $15
\arcsec \times 15 \arcsec $ centered on the red star. North is up and East is left.  
The two lowest panels, {\it c) and d)},
are a further enlargement  of ACS images ($3 \farcs 5\times 3 \farcs 5$)}
\label{f:agb}
\end{figure*}

With a wider baseline it
is possible to separate the sequence of foreground stars and the RGB of SagDIG, 
which is visible in the right panel of Fig. \ref{f:agb}.
\referee{
The foreground stars sequences, which are very narrow in $J-K_s$ color
are much broader in $V-K_s$ and extend up to $V-K_s\sim 7$ (see also Fig. \ref{f:vkjk}).
In the upper right panel of Fig. \ref{f:agb} we show the 
$V-K_s$ color distribution of stars with $20.5<K_s<19.5$.
The detection of the RGB
is clearly visible as an overdensity at  $V-K_s\sim 3.2$.}
The color of RGB stars is consistent with the RGB fiducial lines 
of Milky Way globular clusters taken from 
\cite{vale+2004} of M15 ($\feh=-2.12$, $\mh=-1.91$)) and 
M30 ($\feh=-1.91$, $\mh=-1.71$), confirming
the low metallicity of SagDIG \citep{savi+2002,moma+2005}.
Finally we note that the position of 
the RGB tip is in good agreement
with the value of $K_s=19.5$ expected from the \cite{vale+2004} relations.

\section{SagDIG carbon stars}
In Fig. \ref{f:cmdsag} we show the carbon stars identified
by \cite{cook1987} and \cite{demebatt2002}. Near-IR photometry confirms that six of
them are fainter than the RGB tip ($K_s=19.5$). From our photometry it is
difficult to determine their origin, but we note that the HST
photometry of \cite{moma+2005} demonstrates that three of them have colors
similar to RGB stars, and are likely to belong to an older AGB
population while the others could be Galactic dwarf C-stars.  We
note that there are also four stars detected by \cite{cook1987}
and one from \cite{demebatt2002} that are
brighter than $K_s=19.5$ and bluer than $J-K_s=1.1$, but again in this case
it is impossible to distinguish them from foreground stars.

\referee{ In the upper left panel of Fig. \ref{f:agb} our CMD is
  compared with the main locus of C-stars in LG galaxies from
  \cite{tott+2000}, scaled to the distance of SagDIG.  This C-star
  fiducial suggests that all stars brighter than $K_s=19.5$ and redder
  than $J-K_s=1.1$ are likely SagDIG C-stars.  To minimize
  contamination by unresolved galaxies and stellar blends in our
  selection we assume a magnitude limit of $K_s=18.5$ for C-stars. We
  expect that the red objects at $J-K_s \sim 2$ and fainter than our
  selection, are most likely background galaxies or stellar blends,
  because they are too faint to be C-stars compared to
  \citep{tott+2000} locus.}

Below we describe in detail the 30 candidate C-stars
belonging to SagDIG listed in Table~\ref{t:red}, defined using the 
following criteria: $J-K_s>1.1$ and $K_s<18.5$. 
Firstly we performed a visual
inspection of all objects in the ACS images. The results are reported
in column 6 of Table~\ref{t:red} and in Fig.~\ref{f:agb}.
Four of the C-star candidates fall outside the ACS FOV, one is clearly a background
galaxy, two have shapes that could be associated with irregular background galaxies or
photometric blends. All other objects appear as real stars.
We note that all of our six objects redder than $J-K_s$=2, except for the bright
source with $K_s\simeq 15.5$, are missed by
\cite{cook1987} and \cite{demebatt2002}. 

\referee{We searched for LPV stars comparing ACS $I$-band photometry,
  transformed to standard system according to Eq. (\ref{e:trasfACS}),
  with \cite{demebatt2002} $I$-band photometry of C-star candidates.
  We detected three stars with significant variability, reported in
  the last column of Table \ref{t:red}.  As in the case of DDO~210 our
  candidate LPVs are among the reddest stars, in agreement with the
  color-amplitude relation for LPV stars \citep{raim+2005}.  }

There are also three stars, namely number 6, 7 and 12, with a near-IR
color typical of C-stars, but not identified by \cite{cook1987} or
\cite{demebatt2002}. Suspiciously, all three stars have optical
colors bluer than normal C-stars, i.e. they are the three bluest stars
among our AGB candidates in the right panel of Fig.~\ref{f:agb}, the
only ones with $V-K_s<3.5$.  \referee{The optical colors of these
  three stars are confirmed by the EMMI photometry of
  \cite{moma+2002}.  In the appendix, available in the on-line
  version of this paper, we discuss in further detail the nature of
  these three objects.  Their SED is compatible with a simple model
  obtained summing the SED of a blue and a red star.  The most
  reliable hypothesis is that they are a photometric blends rather
  than binary systems.  The two bluest objects (number 7 and 12) could
  alternatively be high redshift galaxies.  }

As an example, in the lower panels of Fig.~\ref{f:agb} we show a zoom
of SOFI and ACS images in the regions of the two reddest stars of
SagDIG. As can be seen they are very bright in the $K_s$ images and
barely detectable in $J$. Due to the high sensitivity and deep
exposure of the ACS images they are clearly visible in the $I$ frames
even if the flux in the $I$-band is lower than in $J$, but are
extremely faint in $V$. Moreover, especially for star number 9, the
highest sensitivity and spatial resolution of the ACS could resolve in
high detail stars that are blended on the SOFI images. In this
particular case this is not a great problem because the unresolved
stars are extremely faint and do not affect the near-IR measurement
significantly.  Note that we made a visual inspection of all
30 red stars of SagDIG and found that only two suffer from relevant
blending problems.

\subsection{Two-color diagram}
\realfigure{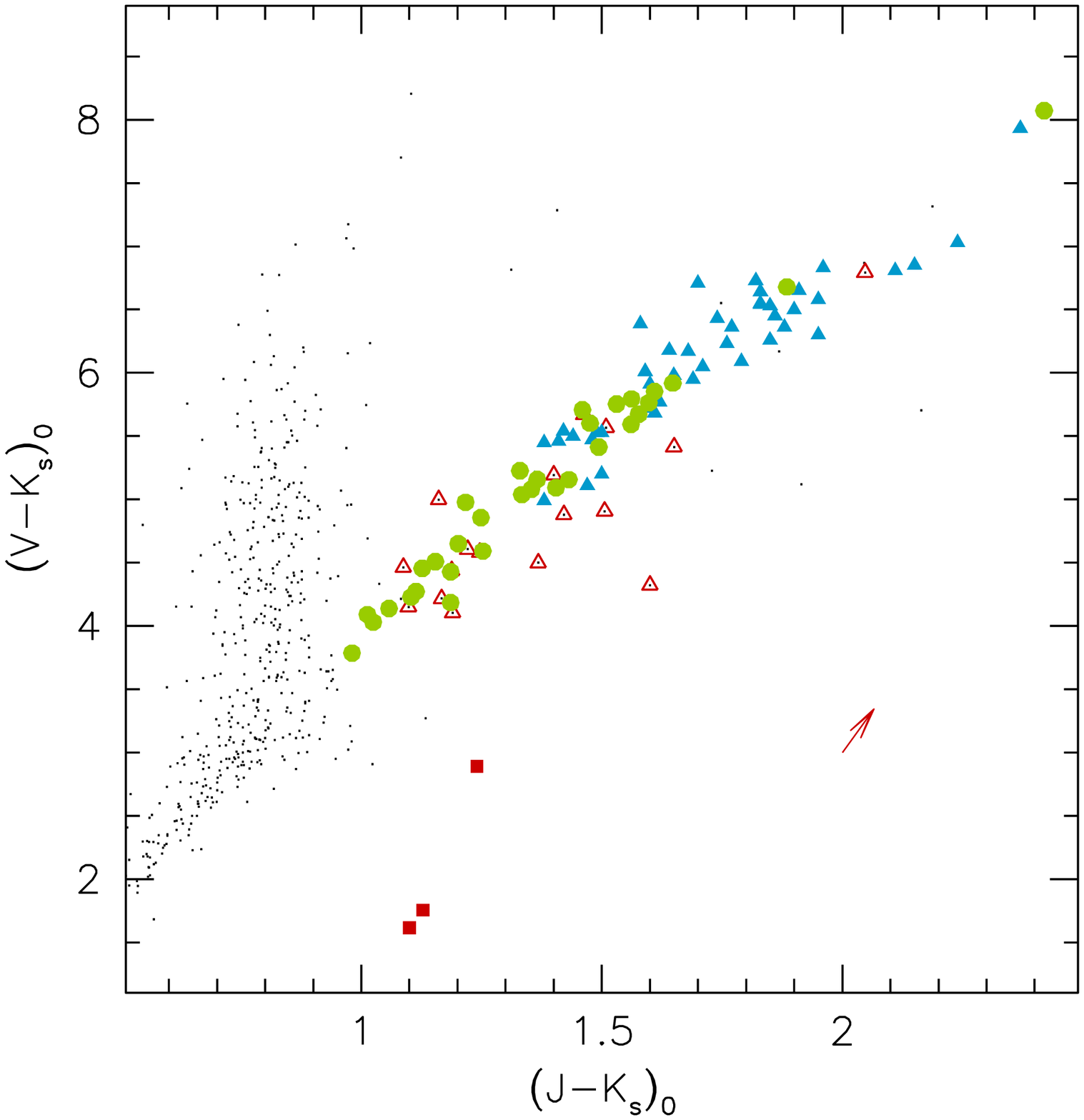}{
Two-color diagram of SagDIG stars (black dots) brighter than $K_s=20$ together with
\cite{berg+2001} Galactic C-stars (large filled blue triangles) and the carbon
stars of Fornax \citep{gull+2006}(large filled green circles). The open red
triangles are the SagDIG C-star candidates, 
\referee{while squares are the three peculiar objects likely not C-stars.}
The arrow represents the reddening
vector toward SagDIG, assuming $E_{(B-V)}=0.12$.}{f:vkjk}

\realfigure{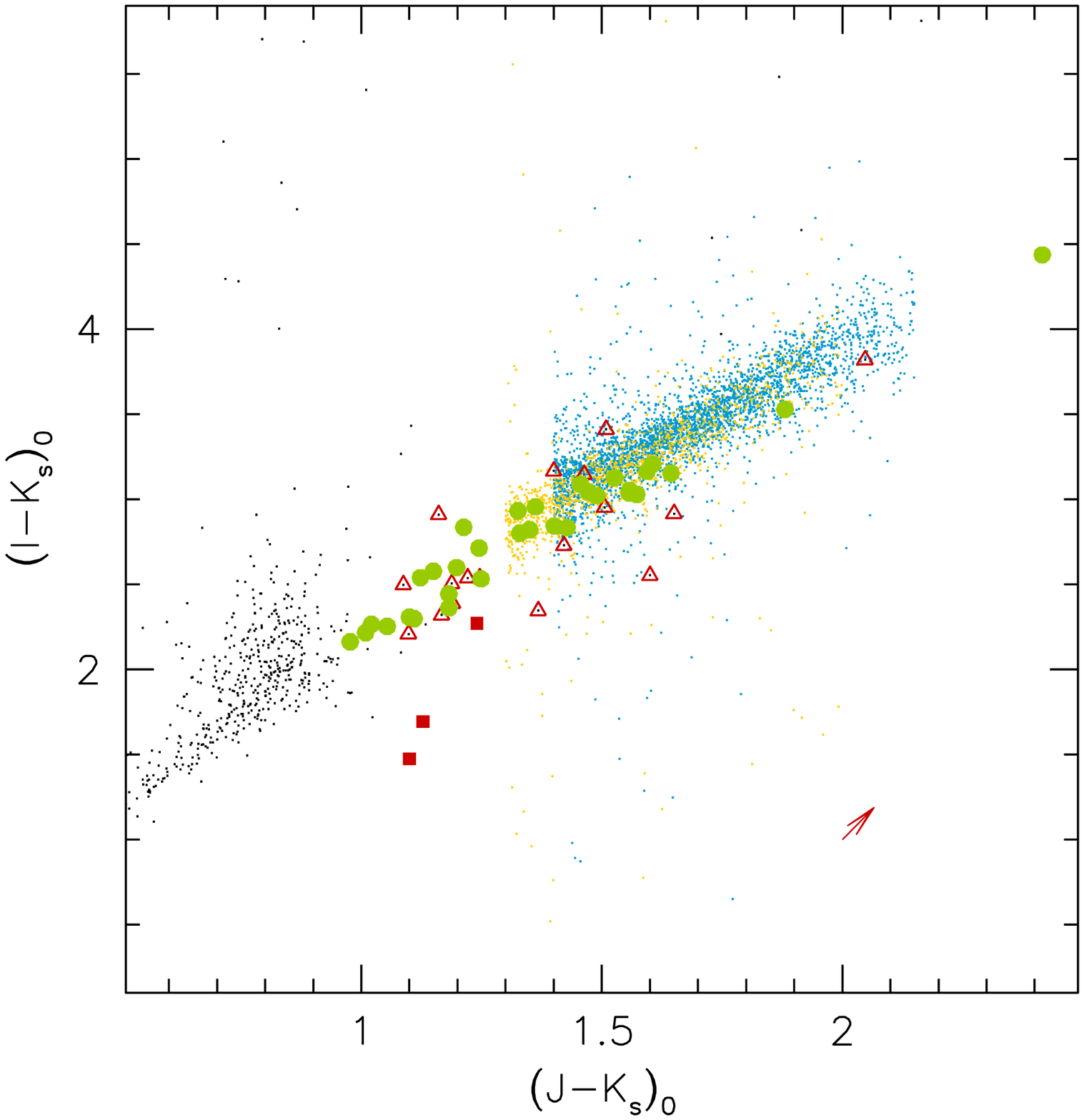}{ Two-color diagram of SagDIG stars brighter
  than $K_s=20$. Small points are C-stars of LMC (blue) and SMC
  (yellow). Filled green circles are C-stars of Fornax.  The open red
  triangles are the SagDIG C-star candidates, \referee{while squares
    are the three peculiar objects likely not C-stars.}  The arrow
  represents the reddening vector toward SagDIG, assuming
  $E_{(B-V)}=0.12$.}{f:ikjk}

In Fig. \ref{f:vkjk} the $(J-K_s)$ vs. $(V-K_s)$ two-color diagram is
shown. The red carbon stars (open triangles) define a quite narrow
sequence, while field stars are well separated and lie on an almost
vertical sequence at $(J-K_s)\simeq0.8$ and $(V-K_s)>3.5$.  On the
same diagram we also show the C-stars of Fornax dSph \citep{gull+2006}
and the galactic C-stars from \cite {berg+2001}.  In Fig. \ref{f:ikjk}
SagDIG stars are shown in the $(J-K_s)$ vs. $(I-K_s)$ diagram,
together with the carbon stars of Fornax \citep{gull+2006} and the
carbon stars of the two Magellanic Clouds extracted from the DENIS
DCMC catalogue \citep{cion+2000}.

The metallicities of the considered systems span a wide range (SagDIG
$\mh \sim -2.0$, SMC $\mh \sim -0.8$, LMC $\mh \sim -0.6$, Fornax $\mh
\sim -0.9$.  Stars from \cite{berg+2001} have a wide range of
metallicity as they are drawn from the Galactic sample), but C-star
sequences from all these systems overlap on a well-defined common
straight line.

\subsection{LF and SFH}
The number of C stars present in SagDIG is sufficient to derive from
their $K_s$ magnitude distribution, an indication about the mean
metallicity and age of the underlying stellar population using the
same criterion adopted by \cite{cion+2006a}.  Probable galaxies and
blends \rereferee{(see note in Table \ref{t:red})} have been excluded
from this analysis.  \rereferee{We note that three objects among our
  C-star candidates (number 6, 7 and 12) have peculiar optical colors,
  and we cannot conclude whether they are C-stars blended with blue
  stars or foreground galaxies. We therefore studied the $K_s$-band LF
  obtained from the whole sample of 27 C-stars candidates, but the
  analysis was repeated excluding the three peculiar objects.  }

The observed $K_s$-band distributions of the C stars is shown in Fig.
\ref{f:histc}.  \rereferee{ When all 27 stars are considered, the LF
  peaks around $K_s=17.3$ and is asymmetric. We note that the three
  peculiar objects populate the peak of the LF, and excluding them
  significantly changes the peak.  }
The two $K_s$-band LFs have been compared with theoretical
distributions created as in \cite{cion+2006a} using up-to-date stellar
evolutionary tracks and a population synthesis code
\citep[TRILEGAL;][]{gira+2005}. The code randomly generates stars
following a given star formation rate (SFR), age-metallicity relation
and initial mass function. The intrinsic stellar properties are
interpolated over a large grid of stellar evolutionary tracks, based
on \cite{bert+1994} and \cite{gira+2003} for massive stars,
\cite{gira+2000} for low- and intermediate-mass stars, and
complemented with the grids of thermally pulsing AGB tracks described
by \cite{mari+1999,mari+2003}.  Near-IR magnitudes were simulated from
bolometric magnitudes and photometric errors were also included. More
details about the construction and properties of the models are given
in \cite{cion+2006a} while the isochrones used in this study are
available at http://pleiadi.oapd.inaf.it. Five different cases for
metallicity and SFR have been adopted resulting in 25 possible
combinations which in turn have been compared with the observed
distribution of C stars. The metallicity values are expressed in terms
of Z$=0.0005$, $0.001$, $0.004$, $0.008$ and $0.016$. The SFR is
described as $\psi(t)\propto\exp(t/\alpha)$, where $t$ is the stellar
age, and $\alpha$ a free parameter. We have considered values for
$\alpha$ of: $-5$, $-2$, $1000$, $5$ and $2$ which correspond to a
mean-age of all stars of $2$, $3.9$, $6.3$, $8.7$ and $10.6$ Gyr
respectively.  The comparison between observed and theoretical
$K_s$-band distributions has been evaluated using the statistical
$\chi ^2$ test where only bins containing more than one source have
been considered.

\realfigure{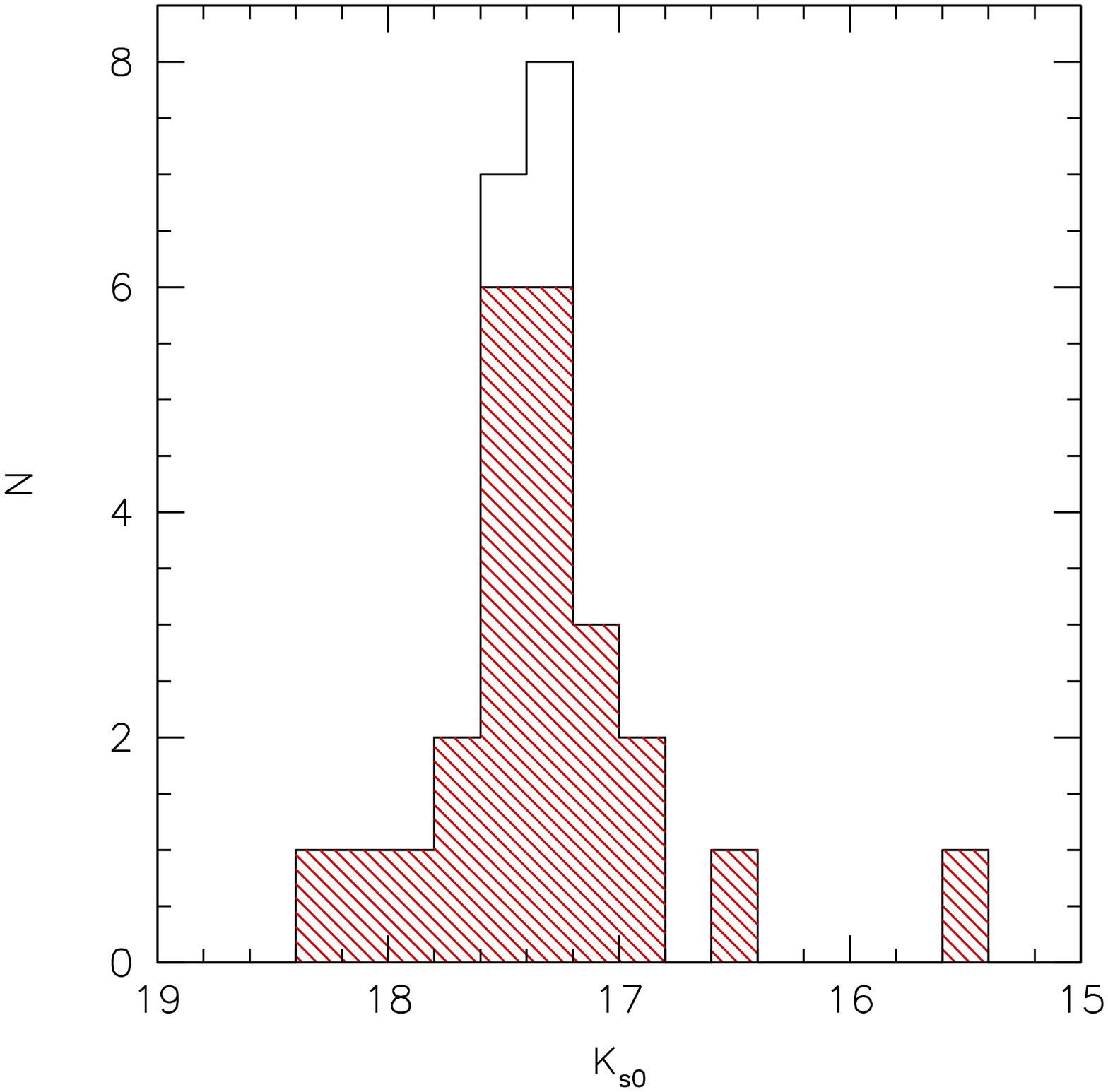}{ Observed $K_{s0}$ de-reddened magnitude
  distribution of C-stars in SagDIG. Bins are of $0.2$ mag.
  \rereferee{The dashed histogram is obtained excluding the three
    peculiar objects with blue optical colors.}  }{f:histc}


\begin{figure}
\includegraphics[width=\columnwidth]{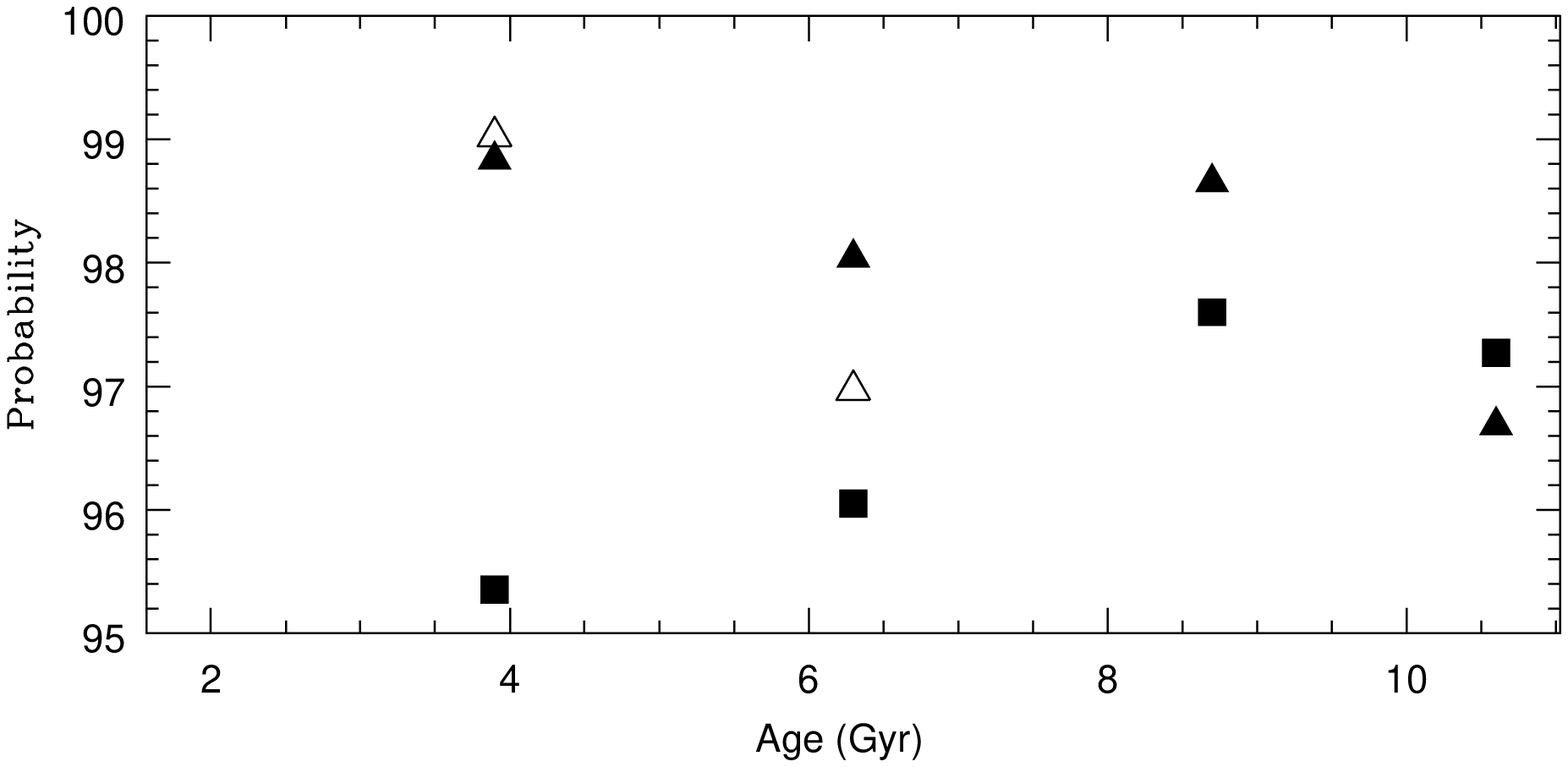}\\
\includegraphics[width=\columnwidth]{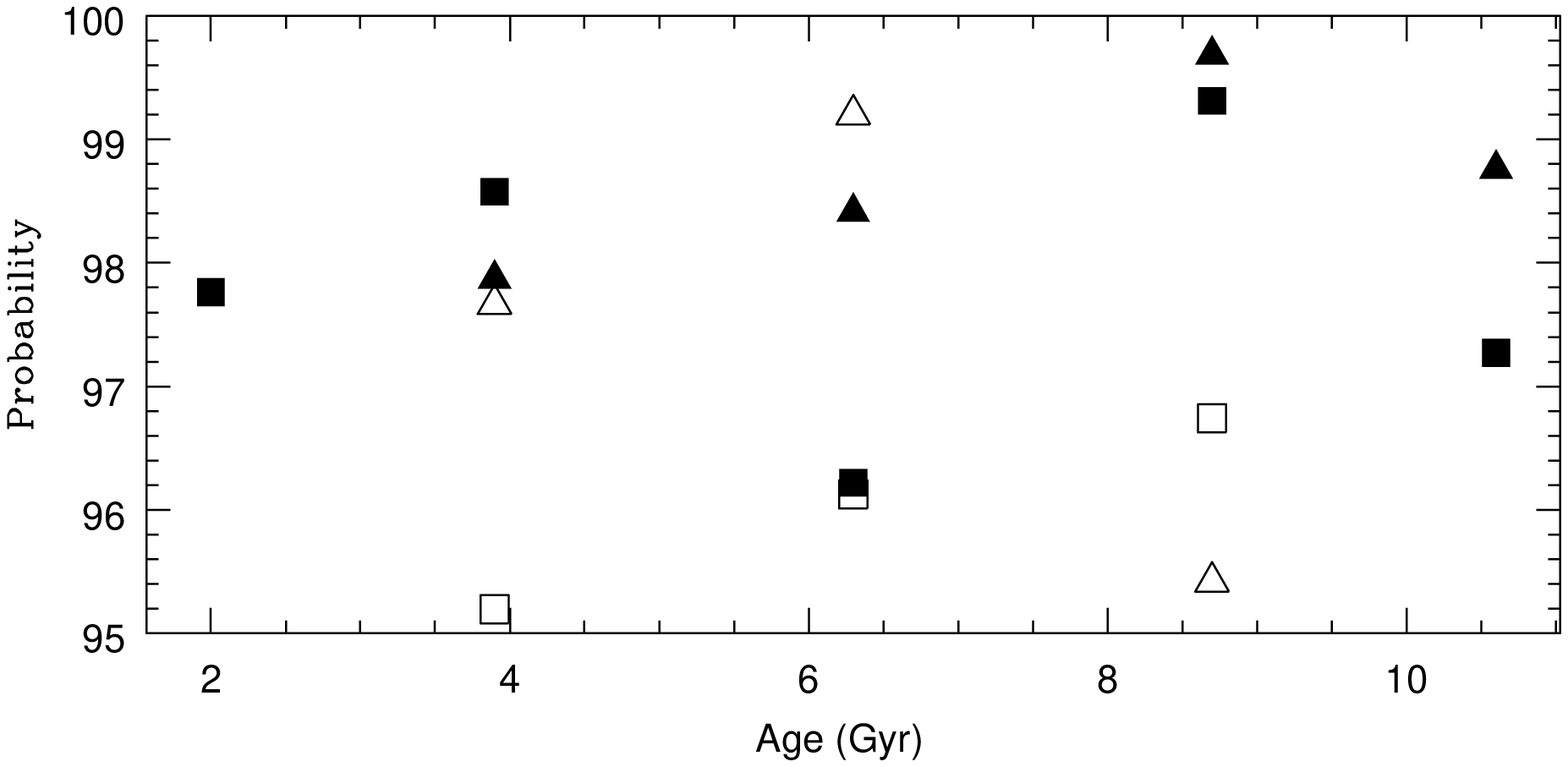}
\caption{
Probability as a function of mean age for models of a given
metallicity which represent the stellar population of SagDIG. Different
symbols refer to a different metallicity as follows: 
Z$=0.0005$ (filled triangles), 
Z$=0.001$ (empty triangles), 
Z$=0.004$ (filled squares),
Z$=0.008$ (empty squares). 
Other models explored in this study give a probability much
lower than the values plotted here and are not shown.
\rereferee{
{\it Upper Panel:} The result obtained considering all the 27
candidate C-stars.  
{\it Lower Panel:} The result obtained excluding 
the three peculiar objects with blue optical colors from the analysis.}
}\label{f:sfh}
\end{figure}

Figure \ref{f:sfh} shows the most probable metallicity at a given age
for the stellar population of SagDIG.  
\rereferee{ In the upper panel we show the probability obtained
  considering all 27 C-star candidates, while the lower panel shows the
  result obtained excluding the three peculiar objects.  The
  point corresponding to the highest probability in the upper panel
  indicates that the overall population is young (on average $4$ Gyr
  old) and metal-poor (at least \feh$<-1.3$ dex assuming
  Z$_{\odot}=0.02$).  However, these numbers need to be taken with
  care due to the limited sample of C-stars and the magnitude binning
  adopted.  When the three peculiar objects are not considered, the
  model with the highest probability turns out to be more metal-poor
  ($Z=0.0005$, corresponding to \feh$=-1.6$).  In order to have an
  estimate of the significance of our results and of the error-bars to
  be associated with the points in Fig. \ref{f:sfh}, we applied the {\sc
    Jacknife} method to our sample.  We repeated our analysis five
  times, removing in turn an object from each of the bins with more
  than one star in the histogram in Fig. \ref{f:histc}.
  This resulted in a large spread in the probabilities corresponding
  to models with $Z>0.0005$. On the other hand for $Z=0.0005$ models,
  a much lower spread ($\simeq 1\%$) is observed for mean ages between
  4 and 8 Gyr.  We therefore conclude that the population of C-stars
  has to be metal-poor, at least $\feh \lesssim -1.3$, and possibly
  $\feh \sim -1.6$.  The age of the overall population cannot be well
  constrained, and we can only conclude that the most probable mean
  age is between 4 and 8 Gyr.  }

SagDIG is considered one of the most, if not the most metal-poor
galaxy in the Local Group.  \cite{moma+2005}, from the color of its
RGB stars, found a metallicity in the range \feh$=-2.1$ to $-1.9$
(depending on the adopted reddening) for the oldest stellar
population.  Their value does not take into account the actual age of
RGB stars of SagDIG, so the calculated value is to be considered a
lower limit, but the authors assert that the correction to be applied
is quite small.  On the other hand, from isochrone fitting of young
blue loop stars, \cite{moma+2005} found a metallicity \feh$\sim -1.7$
for the 100-600 Myr old stars.  From optical spectroscopy of the
brightest \ion{H}{II} region \cite{savi+2002} confirmed that SagDIG is
indeed one of the most metal-poor galaxies.

No detailed studies have been dedicated to the SFH of SagDIG so far.
\cite{kara+1999} photometry was not deep enough to reach the main
sequence turnoff, so they could not resolve the SFH for stars older
than 0.2 Gyr.  The high precision photometric study of
\cite{moma+2005} revealed the presence of a red HB, proving the
presence of an old stellar population, with an age comparable with
that of Galactic globular clusters.  The prominence of the red clump
of He burning stars shows that the dominant stellar population is
older than 1 Gyr.  \cite{moma+2005} showed that the distribution in
the CMD of red clump and main sequence stars is indicative of an
extended period of star formation, spanning from 10 Gyr to $\sim 100$
Myr ago.  The SFH was not derived in detail by \cite{moma+2005}, but a
forthcoming paper will be dedicated to a further investigation with
CMD simulation.

\section{Conclusions}

We investigated the bright AGB content of two Local Group dwarf
irregular galaxies DDO~210 and SagDIG using near-IR and optical
imaging data.

In DDO~210 we detected two out of three previously known C-stars
\citep{battdeme2000}.  The third star is located outside the region
covered by our data.  A further six red objects with $J-K_s>1.1$ and
$K_s<18.5$ have optical and near-IR colors consistent with carbon
giants.  One of these objects, due to its very red color
($J-K_s=2.9$), if confirmed with future higher resolution images
and/or spectroscopy, could be a dust enshrouded AGB star. With only a
handful of C-star candidates it is not possible to put significant
limits on the SFH of this galaxy.  However,
combining these results with much deeper optical imaging may help
constrain the rate of star formation over the last few Gyr \citep[for
recent results see also][]{mcco+2006}.  Similar, dusty AGB stars have
only recently started to be investigated in less distant galaxies
thanks to higher sensitivity of near-IR detectors.
 
SagDIG is a much more luminous galaxy which has experienced a
prolonged star formation \citep{moma+2005}, and consequently it
harbors a rich intermediate-age population. Due to high foreground
contamination, the bluer oxygen-rich AGB stars and bluer C-rich stars
are more easily detectable through the narrow-band filter technique
\citep{cook1987, demebatt2002}. However, the reddest carbon stars and
dust enshrouded AGB stars are easier to pick up through near-IR
imaging. We have combined the results from the literature with our
near-IR images obtained with SOFI at NTT and optical ACS/HST data.
All objects redder than the foreground sequence ($J-K_s>1.1$) have
been carefully inspected and blends or background galaxies were
excluded from the final number of C-stars.  \rereferee{ In the near-IR
  CMD we identified 27 C-stars candidates in SagDIG.  Eighteen of them
  were C-stars previously known from the works of \cite{demebatt2002}
  and \cite{cook1987}.  Six candidate C-stars identified by
  \cite{cook1987} have $K_s$-band magnitudes fainter than the TRGB.
  Four other candidates by \cite{cook1987} and one from
  \cite{demebatt2002} are brighter than the TRGB but cannot be
  distinguished from foreground stars from our near-IR photometry.  }
\rereferee{Among these 27 candidates, three have peculiar optical colors,
  too blue compared to the other C-stars.  We investigated in some
  detail their SEDs concluding that they may be either high-z galaxies
  appearing as point sources in ACS data, or blended stars containing
  blue and red components.}

The LF of the  C-stars in near-IR was  used to investigate the  SFH of
SagDIG  in   the   same way   as was     done  for the  LMC    and SMC
\citep{cion+2006a, cion+2006b}.
\referee{
The result shows that the underlying stellar population of the galaxy
is metal-poor, having at least $\mathrm{[Fe/H]} \la -1.3$, and young,
with the most probable average age between 4 and 8 Gyr
for the dominant population.}

\begin{acknowledgements}
We thank the anonymous referee for many useful suggestions which
improved the presentation.  We gratefully acknowledge G.  Rodighiero
for helpful discussions about the SED, Yazan Momany for his ACS data
and useful comments, Alan McConnachie for providing us with the
electronic version of his photometry.  This publication makes use of
data products from the Two Micron All Sky Survey, which is a joint
project of the University of Massachusetts and the Infrared Processing
and Analysis Center/California Institute of Technology, funded by the
National Aeronautics and Space Administration and the National Science
Foundation.  M.G.\ wishes to thank the European Southern Observatory
for partial funding through DGDF and for hospitality during a visit in
which this paper was partially written.
\end{acknowledgements}

\begin{table*}
  \caption{
    The red stars selected with $J-K_s>1.1$ and $K_s<18.5$. 
Columns 7 and 8 give the identifications from
    \citet[][C87]{cook1987} and \citet[][DB02]{demebatt2002}. The notes in column 6 refer 
    to the classification of the objects based on visual 
    inspection of $V$ and $I$-band HST/ACS images \citep{moma+2005}:
    0: galaxy; 
    1: star;
    2: outside the ACS FOV;
    3: blend. 
    \rereferee{
      Stars marked as {\it Blue} are too blue in optical bands 
     (see text for details).}
    \referee{Last column reports the detection of significant $I$-band variability for
      \cite{demebatt2002} C-star candidates.}
  }
\label{t:red}
\centering
\begin{tabular}{c r@{$^h$} r@{$^m$} l@{\fs} l  r@{\degr} r@{\arcmin} l@{\farcs} l c c l c c r}
\hline\hline
ID&
\multicolumn{4}{c}{$\delta (2000)$}&
\multicolumn{4}{c}{$\alpha (2000)$}&
$K_s$&
$J-K_s$&
\multicolumn{1}{c}{note}&
C87&
DB02&
$\Delta I$\\
\hline
1  & 19 &30 & 08 & 14 &$-17$ &40 &00 & 84 & 17.041 & 1.480 & 2       & 1 & 2 &\\
2  & 19 &30 & 06 & 25 &$-17$ &40 &54 & 66 & 15.616 & 2.360 & 2       & 2 & 5 & \\
3  & 19 &30 & 05 & 17 &$-17$ &41 &42 & 79 & 17.389 & 1.227 & 1       & \nodata & 9 &\\
4  & 19 &30 & 04 & 17 &$-17$ &40 &30 & 27 & 17.754 & 1.153 & 1       & 4 & 8 &\\
5  & 19 &30 & 04 & 08 &$-17$ &40 &15 & 65 & 18.340 & 1.814 & 0       & \nodata & \nodata &\\
6  & 19 &30 & 03 & 25 &$-17$ &41 &33 & 80 & 17.624 & 1.306 & 1,Blue  & \nodata & \nodata &\\
7  & 19 &30 & 02 & 02 &$-17$ &40 &19 & 58 & 17.262 & 1.194 & 1,Blue  & \nodata & \nodata &\\
8  & 19 &30 & 02 & 01 &$-17$ &42 &05 & 77 & 17.036 & 1.487 & 1       & \nodata & 3 &\\
9  & 19 &30 & 01 & 75 &$-17$ &40 &52 & 79 & 17.400 & 3.656 & 1       & \nodata & \nodata &\\
10 & 19 &30 & 01 & 68 &$-17$ &40 &04 & 29 & 17.169 & 1.256 & 1       & 5 & \nodata &\\
11 & 19 &30 & 01 & 67 &$-17$ &41 &06 & 72 & 17.465 & 1.529 & 1       & \nodata & 10 &0.31 \\
12 & 19 &30 & 01 & 12 &$-17$ &39 &59 & 40 & 17.284 & 1.166 & 1,Blue  & \nodata & \nodata &\\
13 & 19 &30 & 01 & 01 &$-17$ &40 &52 & 74 & 17.319 & 1.312 & 1       & 6 & 1 &\\
14 & 19 &30 & 00 & 68 &$-17$ &41 &00 & 69 & 17.017 & 2.113 & 1       & \nodata & \nodata &\\
15 & 19 &30 & 00 & 29 &$-17$ &41 &02 & 92 & 17.107 & 1.716 & 1       & 8 & 15 & $-0.48$\\
16 & 19 &30 & 00 & 23 &$-17$ &40 &54 & 61 & 17.579 & 1.572 & 1       & 9 & 16 &\\
17 & 19 &30 & 00 & 07 &$-17$ &41 &35 & 29 & 17.859 & 1.666 & 1       & \nodata & 4 &0.53\\
18 & 19 &29 & 59 & 79 &$-17$ &41 &22 & 61 & 18.465 & 1.200 & 3       & \nodata & \nodata &\\
19 & 19 &29 & 59 & 77 &$-17$ &41 &05 & 14 & 16.600 & 2.572 & 1       & \nodata & \nodata &\\
20 & 19 &29 & 59 & 76 &$-17$ &40 &34 & 23 & 17.489 & 1.233 & 1       & 11 & \nodata &\\
21 & 19 &29 & 57 & 95 &$-17$ &40 &17 & 54 & 15.578 & 1.397 & 3       & \nodata & \nodata &\\
22 & 19 &29 & 57 & 55 &$-17$ &40 &41 & 96 & 18.128 & 1.164 & 1       & \nodata & 13 &\\
23 & 19 &29 & 56 & 34 &$-17$ &40 &47 & 41 & 17.261 & 1.466 & 1       & \nodata & 11 &\\
24 & 19 &29 & 56 & 08 &$-17$ &41 &21 & 50 & 17.467 & 4.147 & 1       & \nodata & \nodata &\\
25 & 19 &29 & 55 & 24 &$-17$ &40 &22 & 79 & 17.529 & 1.287 & 1       & 16 & 7 &\\
26 & 19 &29 & 55 & 15 &$-17$ &40 &40 & 79 & 17.259 & 1.254 & 1       & 17 & \nodata &\\
27 & 19 &29 & 52 & 90 &$-17$ &40 &32 & 86 & 17.545 & 1.434 & 1       & 19 & 6 &\\
28 & 19 &29 & 52 & 90 &$-17$ &41 &39 & 54 & 17.319 & 1.575 & 1       & \nodata & 14 & \\
29 & 19 &29 & 51 & 64 &$-17$ &39 &44 & 31 & 18.289 & 3.638 & 2       & \nodata & \nodata &\\
30 & 19 &29 & 49 & 21 &$-17$ &41 &03 & 05 & 17.727 & 1.985 & 2       & \nodata & \nodata &\\
\hline
\end{tabular}
\end{table*}

\bibliographystyle{aa} 
\bibliography{sagdig} 

\begin{thebibliography}{53}
\expandafter\ifx\csname natexlab\endcsname\relax\def\natexlab#1{#1}\fi

\bibitem[{{Albert} {et~al.}(2000){Albert}, {Demers}, \& {Kunkel}}]{albe+2000}
{Albert}, L., {Demers}, S., \& {Kunkel}, W.~E. 2000, \aj, 119, 2780

\bibitem[{{Battinelli} \& {Demers}(2000)}]{battdeme2000}
{Battinelli}, P. \& {Demers}, S. 2000, \aj, 120, 1801

\bibitem[{{Battinelli} \& {Demers}(2005)}]{battdeme2005}
{Battinelli}, P. \& {Demers}, S. 2005, \aap, 442, 159

\bibitem[{{Bedin} {et~al.}(2005){Bedin}, {Cassisi}, {Castelli}, {Piotto},
  {Anderson}, {Salaris}, {Momany}, \& {Pietrinferni}}]{bedi+2005}
{Bedin}, L.~R., {Cassisi}, S., {Castelli}, F., {et~al.} 2005, \mnras, 357, 1038

\bibitem[{{Bergeat} {et~al.}(2001){Bergeat}, {Knapik}, \& {Rutily}}]{berg+2001}
{Bergeat}, J., {Knapik}, A., \& {Rutily}, B. 2001, \aap, 369, 178

\bibitem[{{Bertelli} {et~al.}(1994){Bertelli}, {Bressan}, {Chiosi}, {Fagotto},
  \& {Nasi}}]{bert+1994}
{Bertelli}, G., {Bressan}, A., {Chiosi}, C., {Fagotto}, F., \& {Nasi}, E. 1994,
  \aaps, 106, 275

\bibitem[{{Bruzual} \& {Charlot}(2003)}]{bruzchar2003}
{Bruzual}, G. \& {Charlot}, S. 2003, \mnras, 344, 1000

\bibitem[{{Carpenter}(2001)}]{carp2001}
{Carpenter}, J.~M. 2001, \aj, 121, 2851

\bibitem[{{Cioni} {et~al.}(2000){Cioni}, {Loup}, {Habing}, {Fouqu{\'e}},
  {Bertin}, {Deul}, {Egret}, {Alard}, {de Batz}, {Borsenberger}, {Dennefeld},
  {Epchtein}, {Forveille}, {Garz{\'o}n}, {Hron}, {Kimeswenger}, {Lacombe}, {Le
  Bertre}, {Mamon}, {Omont}, {Paturel}, {Persi}, {Robin}, {Rouan}, {Simon},
  {Tiph{\`e}ne}, {Vauglin}, \& {Wagner}}]{cion+2000}
{Cioni}, M.-R., {Loup}, C., {Habing}, H.~J., {et~al.} 2000, \aaps, 144, 235

\bibitem[{{Cioni} {et~al.}(2006{\natexlab{a}}){Cioni}, {Girardi}, {Marigo}, \&
  {Habing}}]{cion+2006a}
{Cioni}, M.-R.~L., {Girardi}, L., {Marigo}, P., \& {Habing}, H.~J.
  2006{\natexlab{a}}, \aap, 448, 77

\bibitem[{{Cioni} {et~al.}(2006{\natexlab{b}}){Cioni}, {Girardi}, {Marigo}, \&
  {Habing}}]{cion+2006b}
{Cioni}, M.-R.~L., {Girardi}, L., {Marigo}, P., \& {Habing}, H.~J.
  2006{\natexlab{b}}, \aap, 452, 195

\bibitem[{{Cioni} \& {Habing}(2005)}]{cionhabi2005}
{Cioni}, M.-R.~L. \& {Habing}, H.~J. 2005, \aap, 429, 837

\bibitem[{{Cioni} {et~al.}(2004){Cioni}, {Habing}, {Loup}, {Epchtein}, \&
  {Deul}}]{cion+2004}
{Cioni}, M.~R.~L., {Habing}, H.~J., {Loup}, C., {Epchtein}, N., \& {Deul}, E.
  2004, The Messenger, 115, 22

\bibitem[{{Cohen} {et~al.}(2003){Cohen}, {Wheaton}, \& {Megeath}}]{cohe+2003}
{Cohen}, M., {Wheaton}, W.~A., \& {Megeath}, S.~T. 2003, \aj, 126, 1090

\bibitem[{{Cook}(1987)}]{cook1987}
{Cook}, K.~H. 1987, Ph.D.~Thesis

\bibitem[{{Cutri} {et~al.}(2003){Cutri}, {Skrutskie}, {van Dyk}, {Beichman},
  {Carpenter}, {Chester}, {Cambresy}, {Evans}, {Fowler}, {Gizis}, {Howard},
  {Huchra}, {Jarrett}, {Kopan}, {Kirkpatrick}, {Light}, {Marsh}, {McCallon},
  {Schneider}, {Stiening}, {Sykes}, {Weinberg}, {Wheaton}, {Wheelock}, \&
  {Zacarias}}]{cutri+2003}
{Cutri}, R.~M., {Skrutskie}, M.~F., {van Dyk}, S., {et~al.} 2003, {2MASS All
  Sky Catalog of point sources.} (The IRSA 2MASS All-Sky Point Source Catalog,
  NASA/IPAC Infrared Science
  Archive.~http://irsa.ipac.caltech.edu/applications/Gator/)

\bibitem[{{Dekker} {et~al.}(1986){Dekker}, {Delabre}, \&
  {Dodorico}}]{dekk+1986}
{Dekker}, H., {Delabre}, B., \& {Dodorico}, S. 1986, in Instrumentation in
  astronomy VI; Proceedings of the Meeting, Tucson, AZ, Mar. 4-8, 1986. Part 1
  (A87-36376 15-35). Bellingham, WA, Society of Photo-Optical Instrumentation
  Engineers, 1986, p. 339-348., ed. D.~L. {Crawford}, 339--348

\bibitem[{{Demers} \& {Battinelli}(2002)}]{demebatt2002}
{Demers}, S. \& {Battinelli}, P. 2002, \aj, 123, 238

\bibitem[{{Demers} {et~al.}(2006){Demers}, {Battinelli}, \&
  {Artigau}}]{deme+2006}
{Demers}, S., {Battinelli}, P., \& {Artigau}, E. 2006, \aap, 456, 905

\bibitem[{{Girardi} {et~al.}(2003){Girardi}, {Bertelli}, {Chiosi}, \&
  {Marigo}}]{gira+2003}
{Girardi}, L., {Bertelli}, G., {Chiosi}, C., \& {Marigo}, P. 2003, in IAU
  Symposium, ed. K.~{van der Hucht}, A.~{Herrero}, \& C.~{Esteban}, 551--+

\bibitem[{{Girardi} {et~al.}(2000){Girardi}, {Bressan}, {Bertelli}, \&
  {Chiosi}}]{gira+2000}
{Girardi}, L., {Bressan}, A., {Bertelli}, G., \& {Chiosi}, C. 2000, \aaps, 141,
  371

\bibitem[{{Girardi} {et~al.}(2005){Girardi}, {Groenewegen}, {Hatziminaoglou},
  \& {da Costa}}]{gira+2005}
{Girardi}, L., {Groenewegen}, M.~A.~T., {Hatziminaoglou}, E., \& {da Costa}, L.
  2005, \aap, 436, 895

\bibitem[{{Groenewegen}(2004)}]{groe2004}
{Groenewegen}, M.~A.~T. 2004, \aap, 425, 595

\bibitem[{{Gullieuszik} {et~al.}(2005){Gullieuszik}, {Held}, {Momany},
  {Saviane}, {Rizzi}, \& {Ortolani}}]{gull+2005}
{Gullieuszik}, M., {Held}, E.~V., {Momany}, Y., {et~al.} 2005, in IAU Colloq.
  198: Near-fields cosmology with dwarf elliptical galaxies, ed. H.~{Jerjen} \&
  B.~{Binggeli}, 47--48

\bibitem[{{Gullieuszik} {et~al.}(2007){Gullieuszik}, {Held}, {Rizzi},
  {Saviane}, {Momany}, \& {Ortolani}}]{gull+2006}
{Gullieuszik}, M., {Held}, E.~V., {Rizzi}, L., {et~al.} 2007, \aap, 467, 1025

\bibitem[{{Kang} {et~al.}(2005){Kang}, {Sohn}, {Rhee}, {Shin}, {Chun}, \&
  {Kim}}]{kang+2005}
{Kang}, A., {Sohn}, Y.-J., {Rhee}, J., {et~al.} 2005, \aap, 437, 61

\bibitem[{{Karachentsev} {et~al.}(1999){Karachentsev}, {Aparicio}, \&
  {Makarova}}]{kara+1999}
{Karachentsev}, I., {Aparicio}, A., \& {Makarova}, L. 1999, \aap, 352, 363

\bibitem[{{Lan{\c c}on} \& {Mouhcine}(2002)}]{lancmouc2002}
{Lan{\c c}on}, A. \& {Mouhcine}, M. 2002, \aap, 393, 167

\bibitem[{{Lee} {et~al.}(1999){Lee}, {Aparicio}, {Tikonov}, {Byun}, \&
  {Kim}}]{lee+1999}
{Lee}, M.~G., {Aparicio}, A., {Tikonov}, N., {Byun}, Y.-I., \& {Kim}, E. 1999,
  \aj, 118, 853

\bibitem[{{Lo} {et~al.}(1993){Lo}, {Sargent}, \& {Young}}]{lo+1993}
{Lo}, K.~Y., {Sargent}, W.~L.~W., \& {Young}, K. 1993, \aj, 106, 507

\bibitem[{{Maraston} {et~al.}(2006){Maraston}, {Daddi}, {Renzini}, {Cimatti},
  {Dickinson}, {Papovich}, {Pasquali}, \& {Pirzkal}}]{mara+2006}
{Maraston}, C., {Daddi}, E., {Renzini}, A., {et~al.} 2006, \apj, 652, 85

\bibitem[{{Marigo} {et~al.}(1999){Marigo}, {Girardi}, \& {Bressan}}]{mari+1999}
{Marigo}, P., {Girardi}, L., \& {Bressan}, A. 1999, \aap, 344, 123

\bibitem[{{Marigo} {et~al.}(2003){Marigo}, {Girardi}, \& {Chiosi}}]{mari+2003}
{Marigo}, P., {Girardi}, L., \& {Chiosi}, C. 2003, \aap, 403, 225

\bibitem[{{McConnachie} {et~al.}(2006){McConnachie}, {Arimoto}, {Irwin}, \&
  {Tolstoy}}]{mcco+2006}
{McConnachie}, A.~W., {Arimoto}, N., {Irwin}, M., \& {Tolstoy}, E. 2006,
  \mnras, 373, 715

\bibitem[{{Momany} {et~al.}(2005){Momany}, {Held}, {Saviane}, {Bedin},
  {Gullieuszik}, {Clemens}, {Rizzi}, {Rich}, \& {Kuijken}}]{moma+2005}
{Momany}, Y., {Held}, E.~V., {Saviane}, I., {et~al.} 2005, \aap, 439, 111

\bibitem[{{Momany} {et~al.}(2002){Momany}, {Held}, {Saviane}, \&
  {Rizzi}}]{moma+2002}
{Momany}, Y., {Held}, E.~V., {Saviane}, I., \& {Rizzi}, L. 2002, \aap, 384, 393

\bibitem[{{Moorwood} {et~al.}(1998){Moorwood}, {Cuby}, \& {Lidman}}]{moor+1998}
{Moorwood}, A., {Cuby}, J.-G., \& {Lidman}, C. 1998, The Messenger, 91, 9

\bibitem[{{Nikolaev} \& {Weinberg}(2000)}]{nikowein2000}
{Nikolaev}, S. \& {Weinberg}, M.~D. 2000, \apj, 542, 804

\bibitem[{{Nowotny} {et~al.}(2001){Nowotny}, {Kerschbaum}, {Schwarz}, \&
  {Olofsson}}]{nowo+2001}
{Nowotny}, W., {Kerschbaum}, F., {Schwarz}, H.~E., \& {Olofsson}, H. 2001,
  \aap, 367, 557

\bibitem[{{Persson} {et~al.}(1998){Persson}, {Murphy}, {Krzeminski}, {Roth}, \&
  {Rieke}}]{pers+1998}
{Persson}, S.~E., {Murphy}, D.~C., {Krzeminski}, W., {Roth}, M., \& {Rieke},
  M.~J. 1998, \aj, 116, 2475

\bibitem[{{Raimondo} {et~al.}(2005){Raimondo}, {Cioni}, {Rejkuba}, \&
  {Silva}}]{raim+2005}
{Raimondo}, G., {Cioni}, M.-R.~L., {Rejkuba}, M., \& {Silva}, D.~R. 2005, \aap,
  438, 521

\bibitem[{{Rejkuba} {et~al.}(2005){Rejkuba}, {Greggio}, {Harris}, {Harris}, \&
  {Peng}}]{rejk+2005}
{Rejkuba}, M., {Greggio}, L., {Harris}, W.~E., {Harris}, G.~L.~H., \& {Peng},
  E.~W. 2005, \apj, 631, 262

\bibitem[{{Rejkuba} {et~al.}(2001){Rejkuba}, {Minniti}, {Silva}, \&
  {Bedding}}]{rejk+2001}
{Rejkuba}, M., {Minniti}, D., {Silva}, D.~R., \& {Bedding}, T.~R. 2001, \aap,
  379, 781

\bibitem[{{Renzini} \& {Buzzoni}(1986)}]{renzbuzz1986}
{Renzini}, A. \& {Buzzoni}, A. 1986, in ASSL Vol. 122: Spectral Evolution of
  Galaxies, ed. C.~{Chiosi} \& A.~{Renzini}, 195--231

\bibitem[{{Rieke} \& {Lebofsky}(1985)}]{rieklebo1985}
{Rieke}, G.~H. \& {Lebofsky}, M.~J. 1985, \apj, 288, 618

\bibitem[{{Salaris} {et~al.}(1993){Salaris}, {Chieffi}, \&
  {Straniero}}]{sala+1993}
{Salaris}, M., {Chieffi}, A., \& {Straniero}, O. 1993, \apj, 414, 580

\bibitem[{{Saviane} {et~al.}(2002){Saviane}, {Rizzi}, {Held}, {Bresolin}, \&
  {Momany}}]{savi+2002}
{Saviane}, I., {Rizzi}, L., {Held}, E.~V., {Bresolin}, F., \& {Momany}, Y.
  2002, \aap, 390, 59

\bibitem[{{Stetson}(1987)}]{stet1987}
{Stetson}, P.~B. 1987, \pasp, 99, 191

\bibitem[{{Stetson}(1993)}]{stet1993}
{Stetson}, P.~B. 1993, in IAU Colloq. 136: Stellar Photometry - Current
  Techniques and Future Developments, ed. C.~J. {Butler} \& I.~{Elliott},
  291--+

\bibitem[{{Stetson}(1994)}]{stet1994}
{Stetson}, P.~B. 1994, \pasp, 106, 250

\bibitem[{{Stetson}(2000)}]{stet2000}
{Stetson}, P.~B. 2000, \pasp, 112, 925

\bibitem[{{Totten} {et~al.}(2000){Totten}, {Irwin}, \& {Whitelock}}]{tott+2000}
{Totten}, E.~J., {Irwin}, M.~J., \& {Whitelock}, P.~A. 2000, \mnras, 314, 630

\bibitem[{{Valenti} {et~al.}(2004){Valenti}, {Ferraro}, \&
  {Origlia}}]{vale+2004}
{Valenti}, E., {Ferraro}, F.~R., \& {Origlia}, L. 2004, \mnras, 354, 815

\end{thebibliography}

\Online 
\begin{appendix}
\section{The SED of the three peculiar objects}

\begin{figure*}
\centering
\begin{tabular}{c@{\hspace{5mm}}c@{\hspace{5mm}}c}
Object n$^\circ$ 6&
Object n$^\circ$ 7&
Object n$^\circ$ 12\\
\includegraphics[width=5.5 cm]{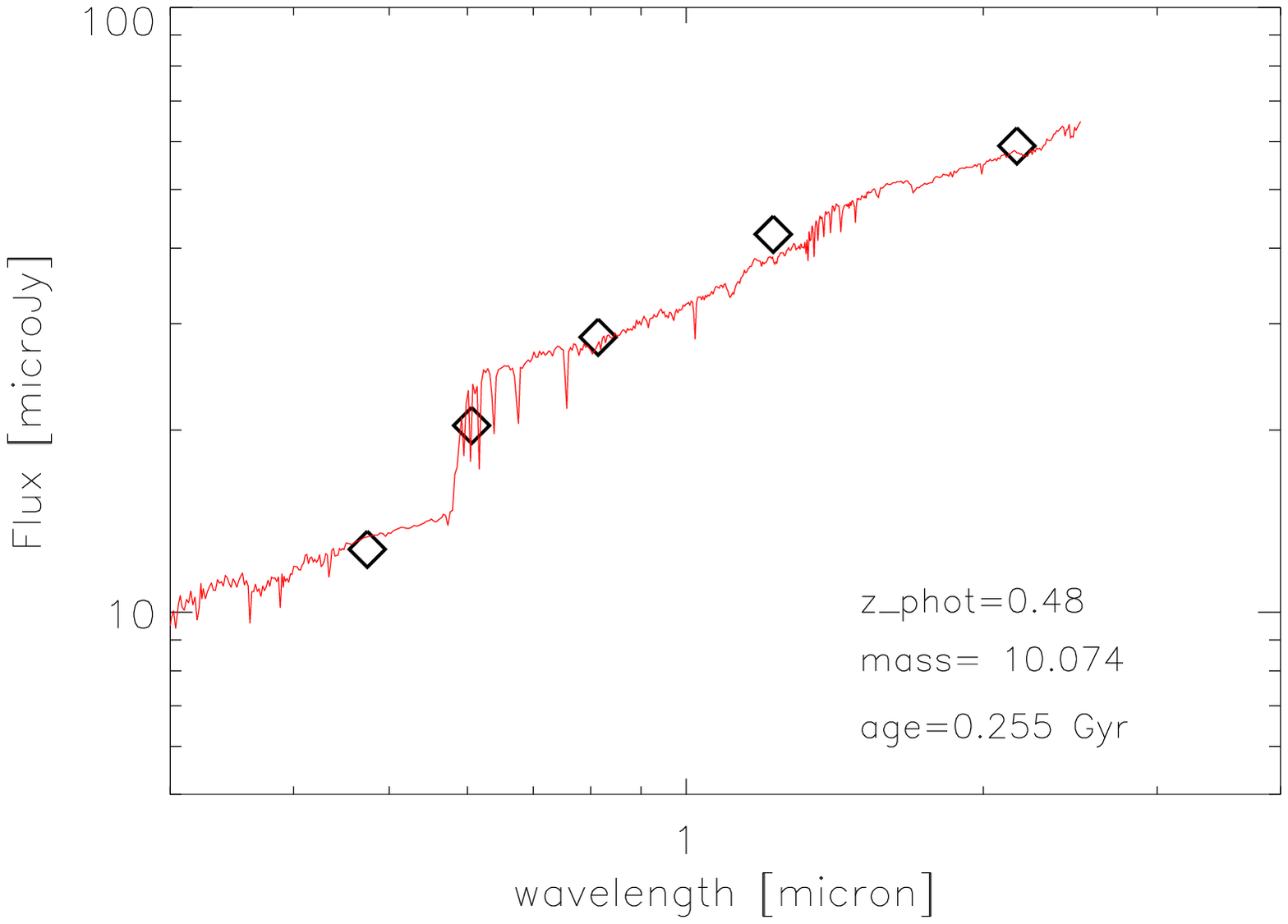}&
\includegraphics[width=5.5 cm]{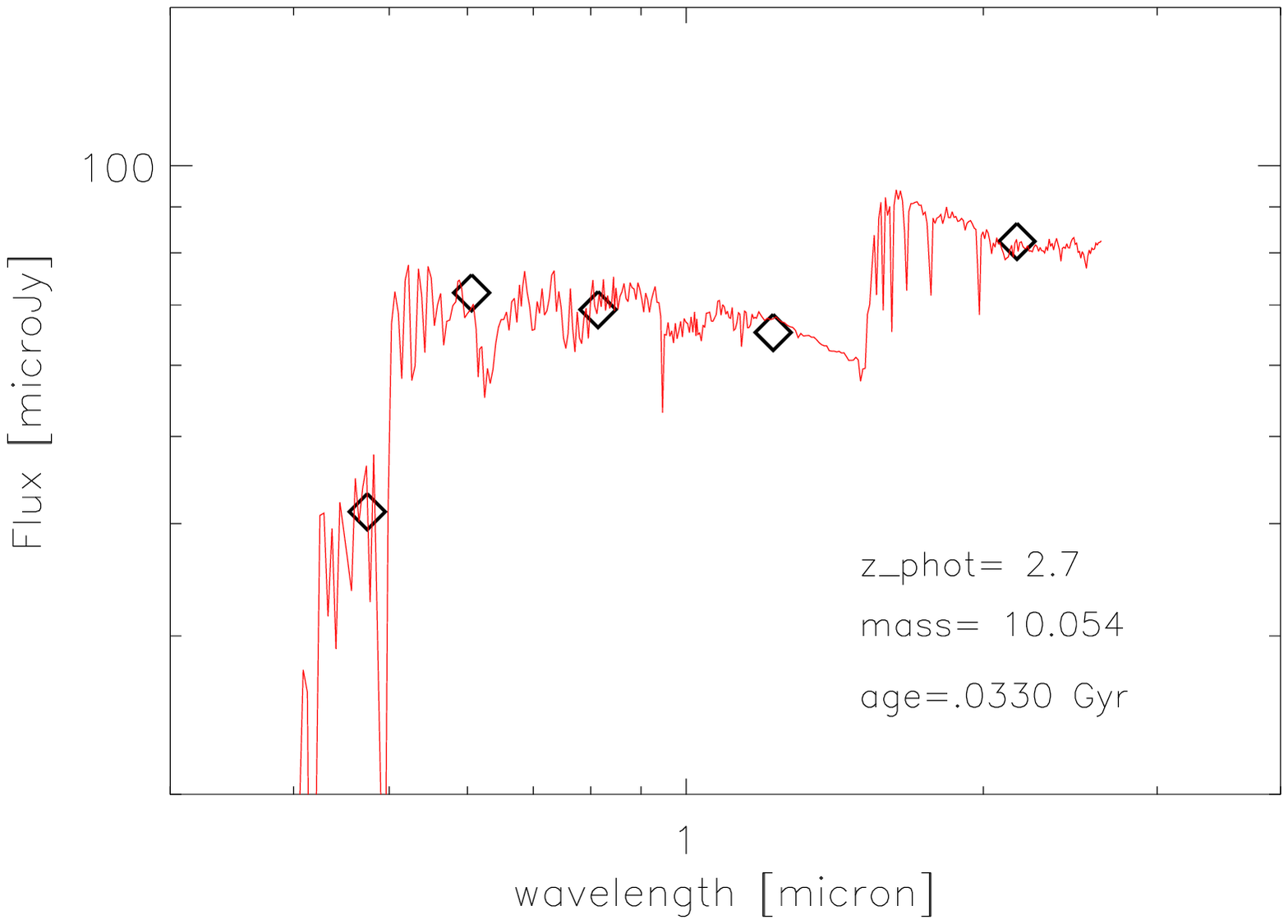}&
\includegraphics[width=5.5 cm]{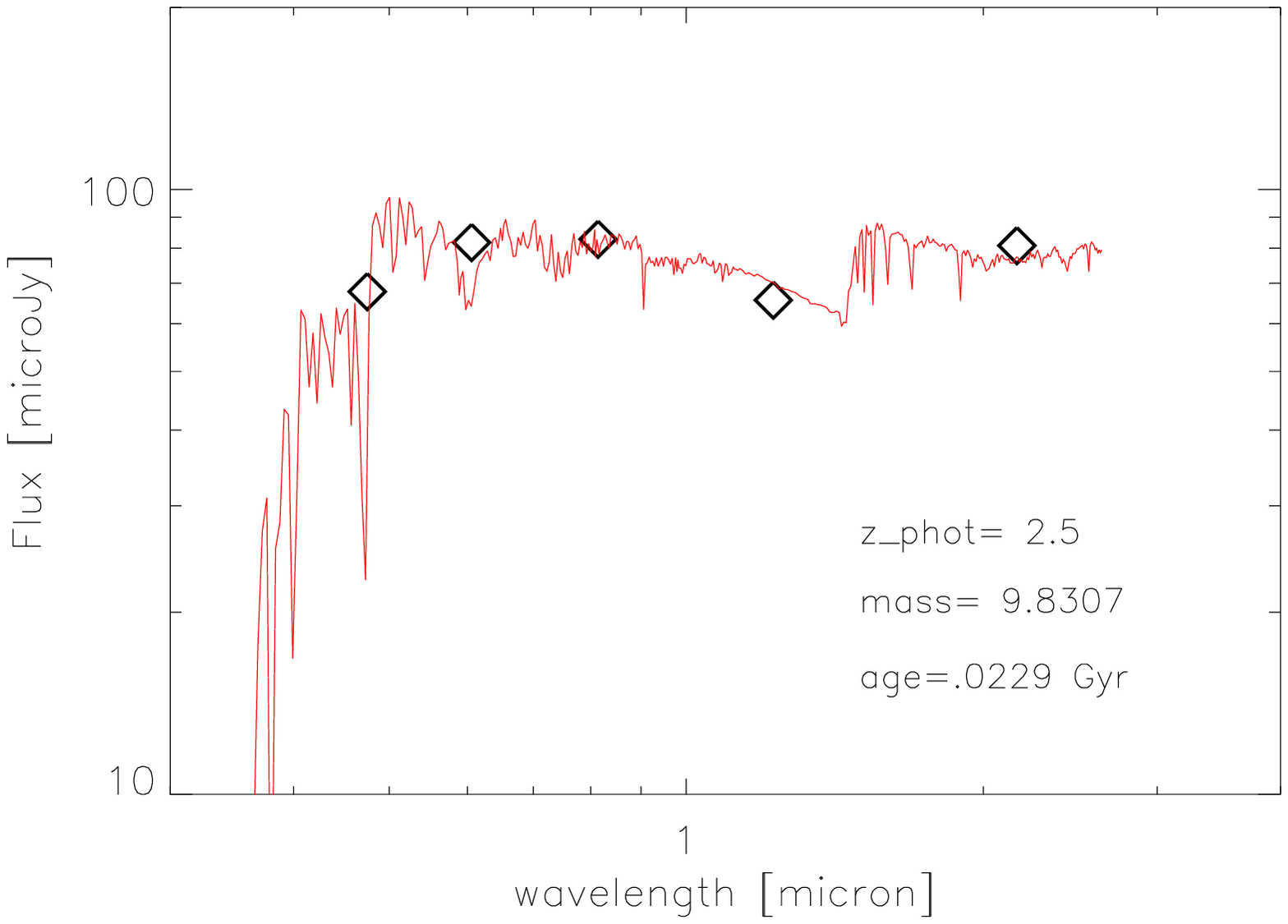}\\
\includegraphics[width=5.5 cm]{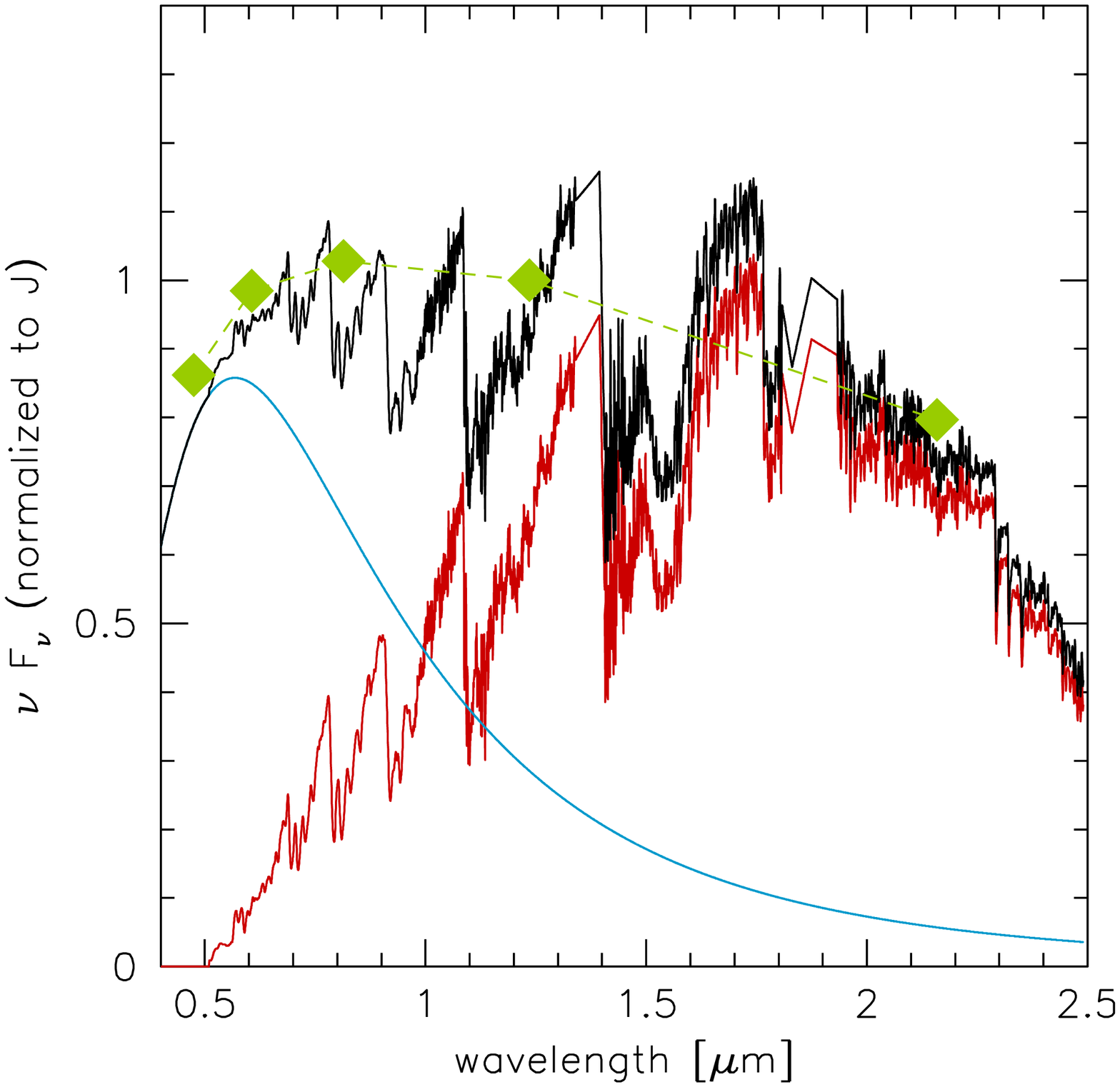}&
\includegraphics[width=5.5 cm]{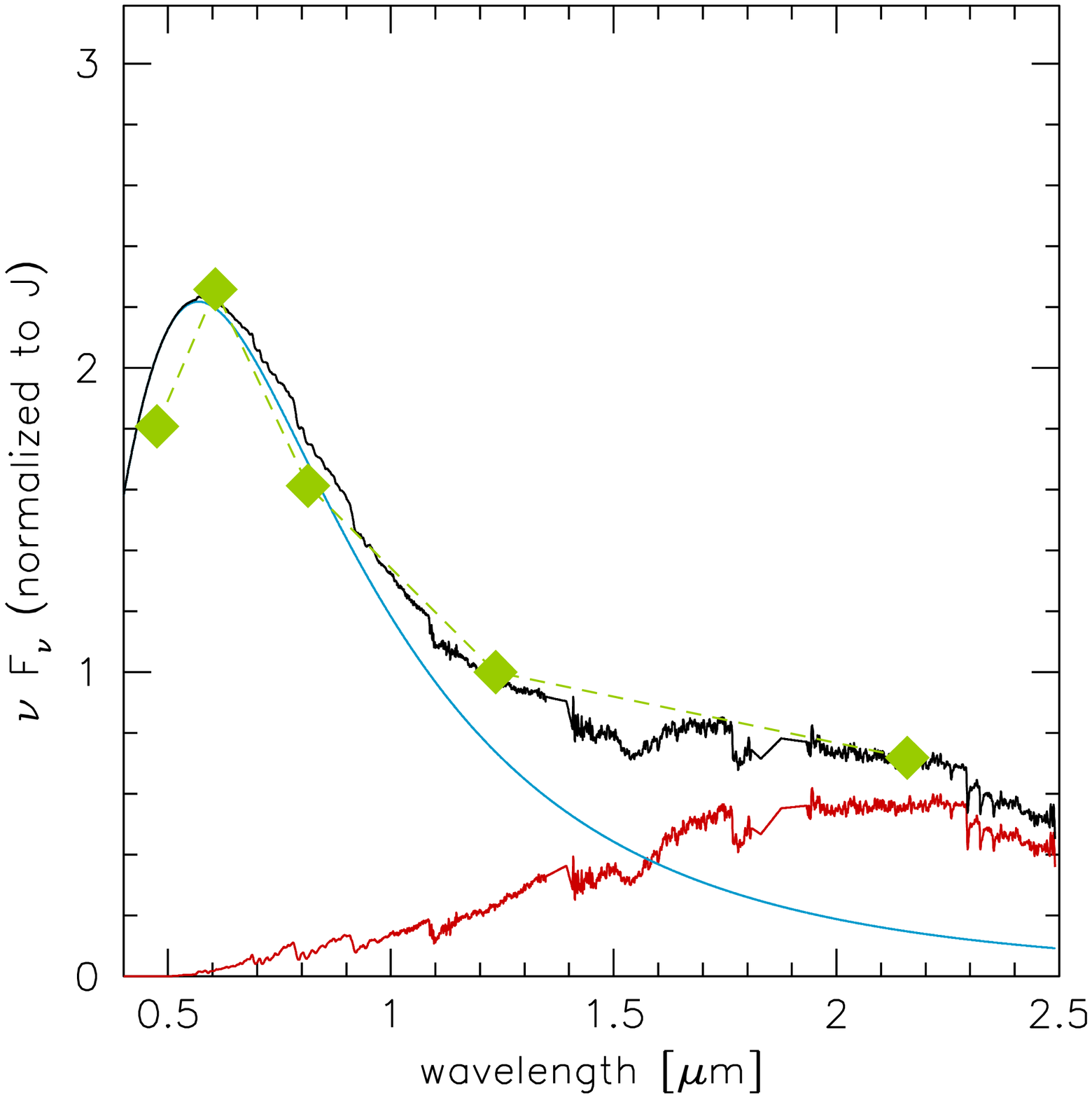}&
\includegraphics[width=5.5 cm]{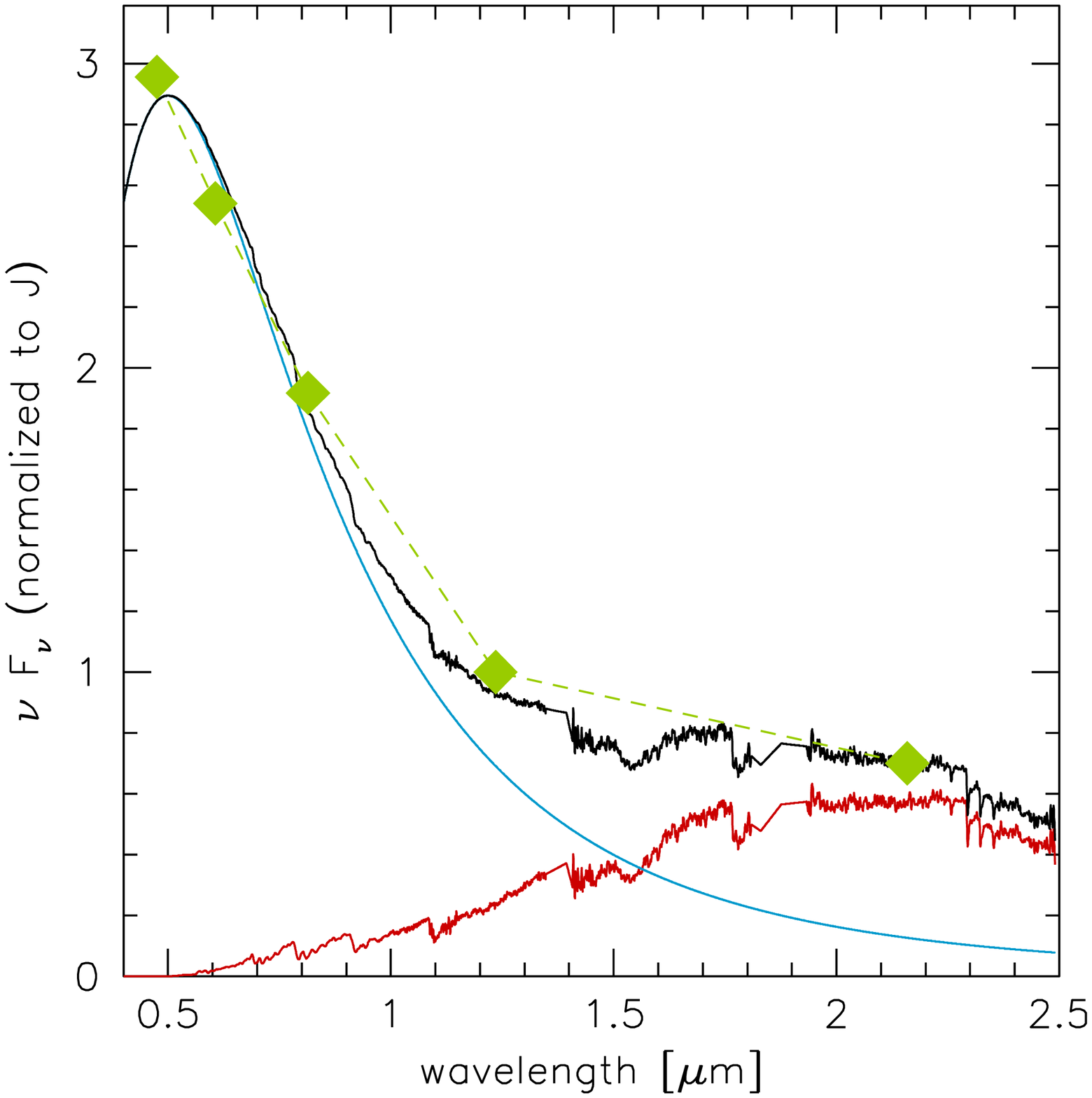}\\
\includegraphics[width=5.5 cm]{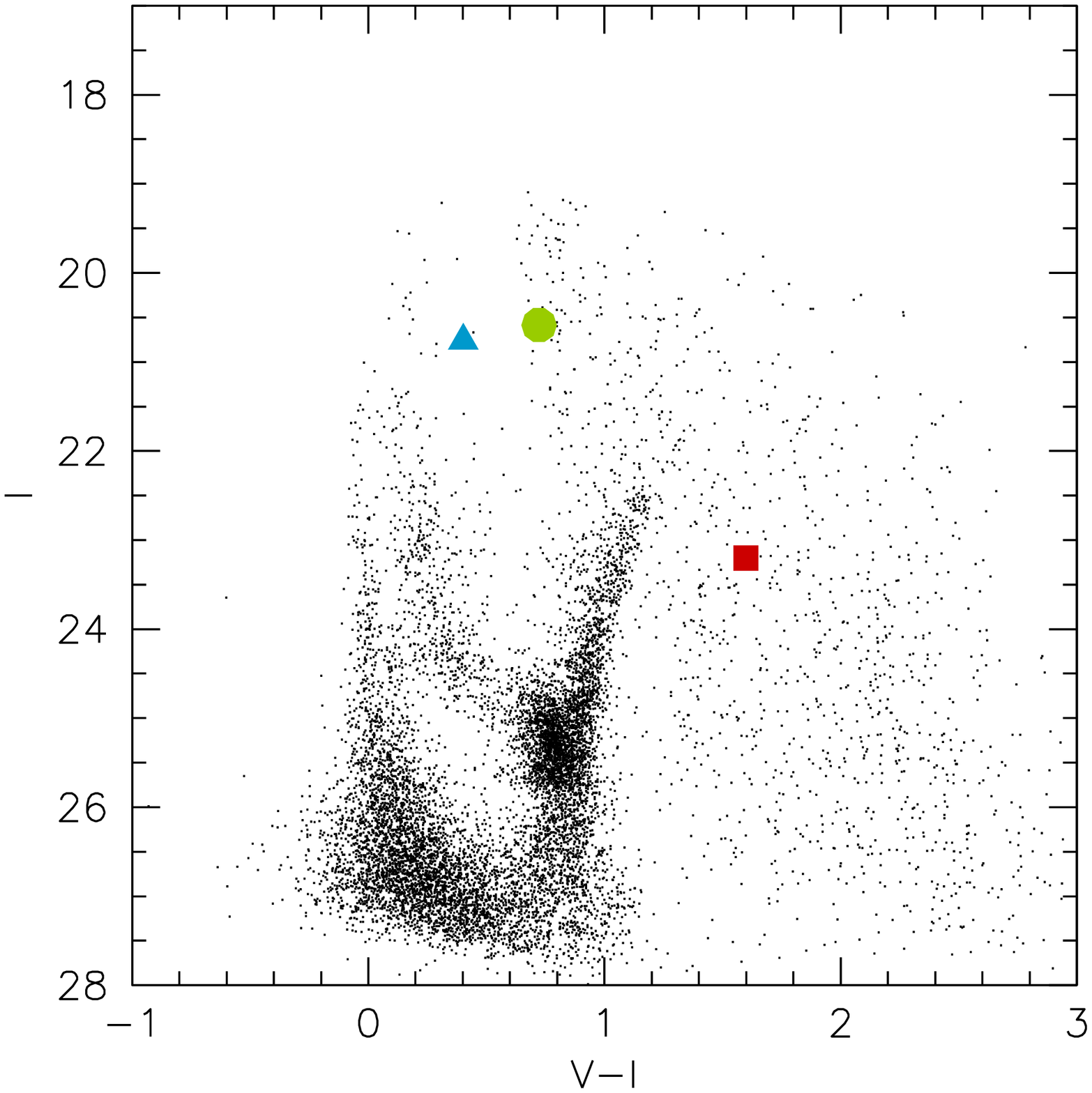}&
\includegraphics[width=5.5 cm]{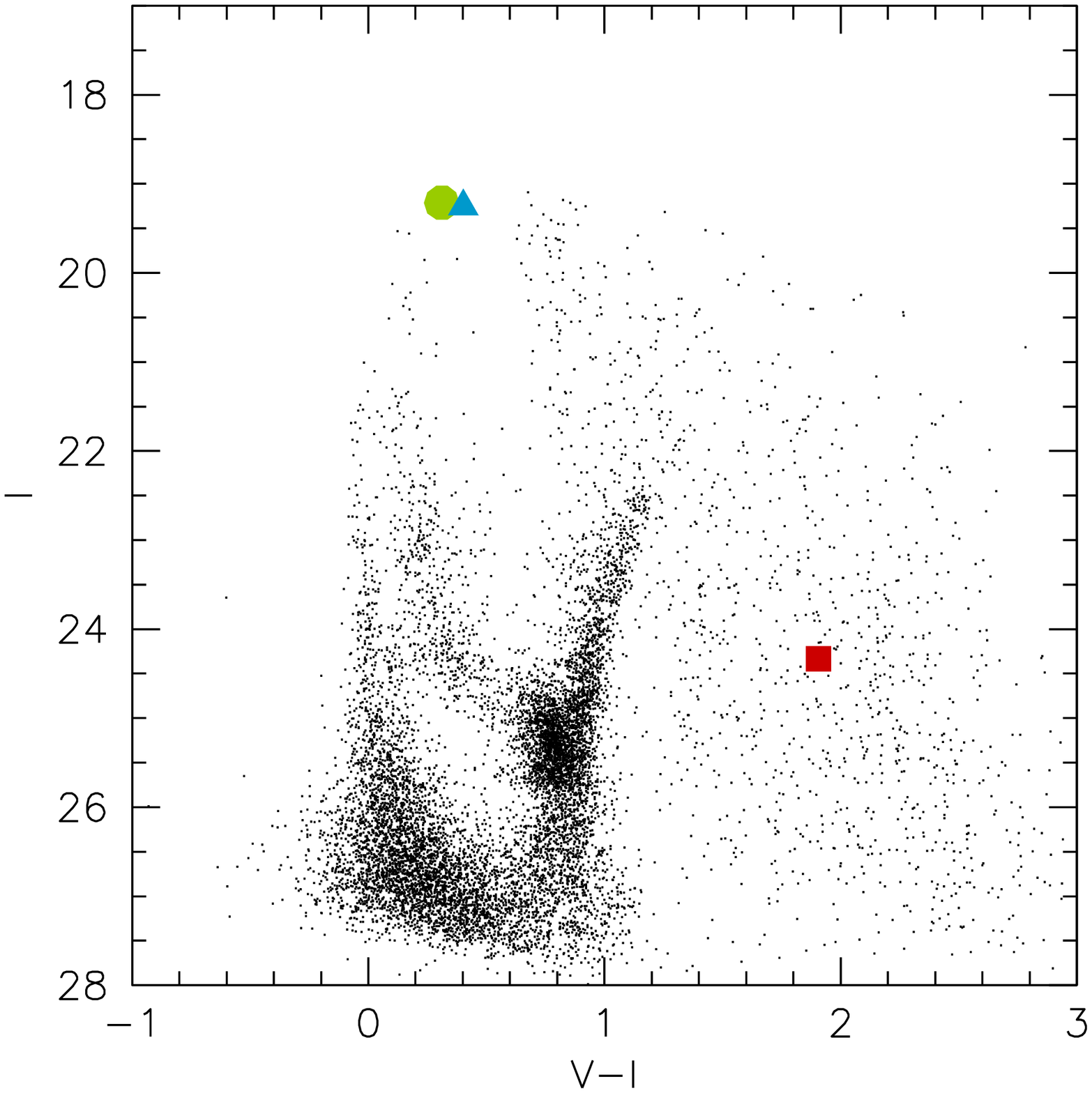}&
\includegraphics[width=5.5 cm]{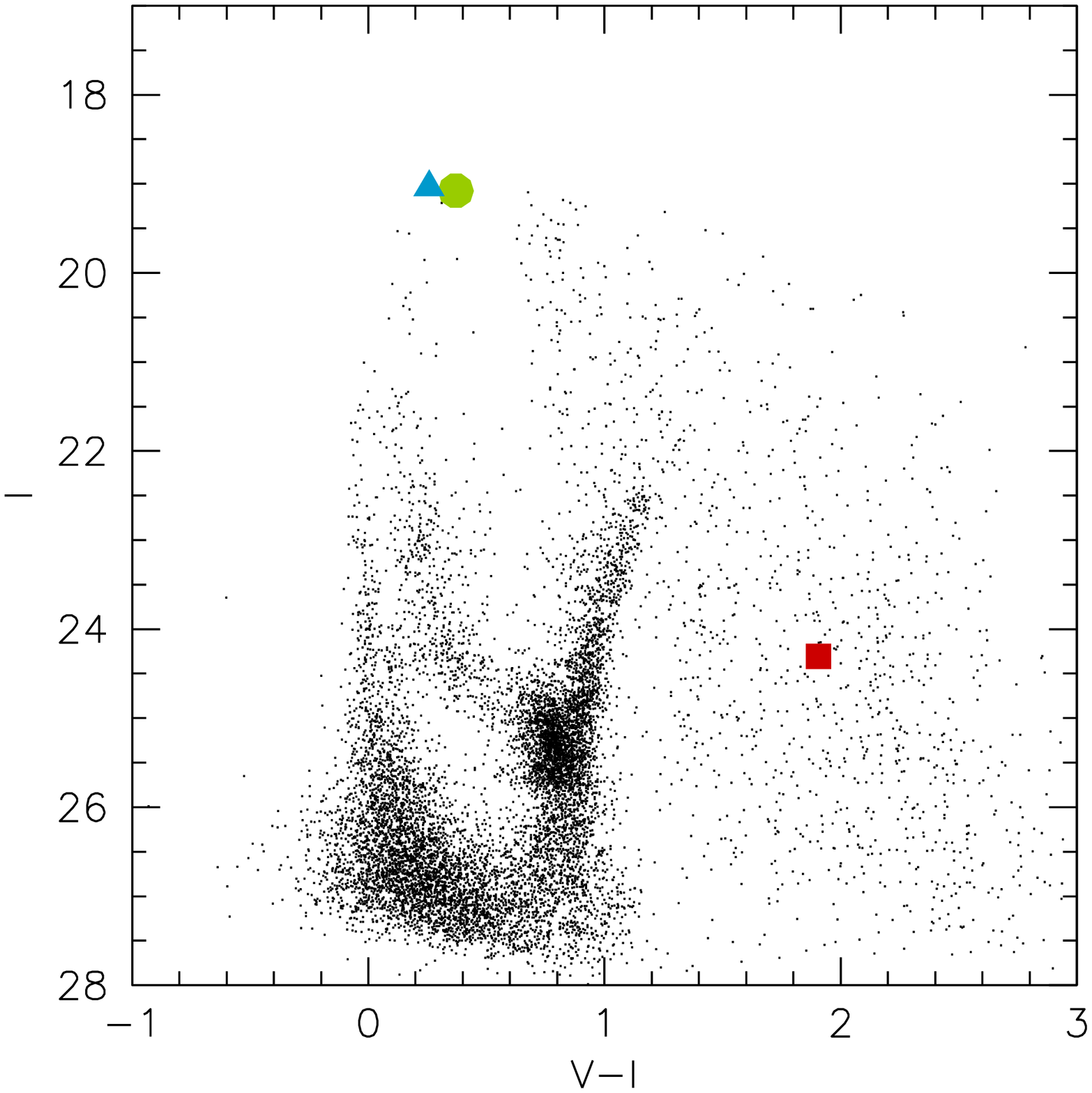}\\
\end{tabular}
\caption{
The SED of objects number 6, 7 and 12.
{\it Upper Panels:} The flux 
together with the best fit models of distant galaxies. 
{\it Central Panels:} The flux density together with 
the {\it double star} models: the SED is fitted by the sum 
of a carbon star spectrum taken from \cite{lancmouc2002} 
and a black body. 
{\it Lower Panels:} The ACS CMD of SagDIG and the photometric points
calculated from the black body model (triangles) and the carbon star
spectra (squares). The measured magnitudes of the real objects are
plotted as circles.
}\label{f:sedGAL}
\end{figure*}

We point out the strange colors of the three red stars (number 6, 7 and
12) with near-IR colors typical of C-stars but not classified as
C-stars by previous studies.  In the $(J-K_s) \ vs.\ (V-K_s)$ diagram
of Fig. \ref{f:vkjk} they are the only stars with $V-K_s<3.5$ (the
other star is the blue object classified as blended, see Fig.
\ref{f:agb}).  In the $(J-K_s) \ vs.\ (I-K_s)$ diagram of Fig.
\ref{f:ikjk} only two of them have blue optical colors, bluer than
$I-K_s=2$. The third one (number 6) is placed among ``normal'' C-stars, 
at $I-K_s=2.24$.  We can therefore say that these three stars show  
SEDs which are similar to a C-star SED in the near-IR, but have a {\it
  blue excess}, which is less pronounced for star number 6.  We
explore the following possible explanations: i) they are high-z
galaxies with a distance sufficient to make them appear as point-like
sources and producing a SED that could explain their color; ii) they
are double stars composed of a very red and a very blue star.  The
latter hypothesis could be due to a blend of two stars or an
intrinsic binary.

We derived the SEDs of these three objects using the zero point taken
from \cite{bedi+2005} for the ACS photometry and from \cite{cohe+2003}
for the near-IR.  First we tested the high redshift galaxy hypothesis,
reproducing the observed SEDs with templates generated with the
stellar population synthesis code {\sc Galaxev} \citep{bruzchar2003}.
Our best fit models are shown in the upper panels of
Fig.~\ref{f:sedGAL}.  The SEDs are compatible with a young
$\sim10^{10} M_\odot$ galaxy at redshift $z\simeq2.5$ for objects
number 7 and 12, and $z\simeq0.5$ for object number 6.  In a standard
cosmological model ($H_0 = 71$, $\Omega_M=0.3$, $\Omega_\Lambda=0.7$)
$z = 2.5$ corresponds to a luminosity distance $D_L \simeq 20000$ Mpc.
The ACS scale is $0\farcs05$ pixel$^{-1}$, hence a pixel corresponds
to $\sim5$ kpc for an object located at 20000 Mpc.  Given that the
typical PSF is 4 pixels, a point-like source located at $z=2.5$ must
have a $FWHM<20$ kpc. This is reasonable for a $10^{10}M_\odot$ galaxy
in which we actually observe only the youngest stars forming possibly
in the central regions observed in the rest-frame ultra-violet with
ACS.  We therefore conclude that two objects (number 7 and 12) could
be galaxies at $z\simeq2.5$.  The third one, at $z\simeq0.5$, should
be located at a distance of an order of magnitude smaller and thus
should be one order of magnitude smaller, in order to appear as a
point like source.  Such a small distance is incompatible with a
$10^{10}M_\odot$ galaxy.  \referee{ The purpose of the discussion is
  to show the viability of the galaxy hypothesis, and hence a
  detailed discussion of other possible solutions within the range of
  explored parameters is beyond the scope of this paper.}

Another possible explanation is that they are double stars or a pair
of unresolved stars, composed of a red and a blue star. The resultant
sum of their SEDs would have blue optical colors and red near-IR
colors. To test this hypothesis, we used the spectral library of
C-stars from \cite{lancmouc2002}.  We added to a C-star the SED of a
simple black body. We consider this a good approximation of a
moderately hot star, sufficient for the qualitative aim of this
analysis.  In the central panels of Fig.~\ref{f:sedGAL} the resulting
fits are presented and appear to be reasonably good.  We derived the
magnitudes of the two components of the best fit models. In the lower 
panels of Fig.~\ref{f:sedGAL} the optical magnitudes of the C-star and
the black body are shown in the ACS CMD of SagDIG.

All three objects are compatible with a SED composed of
a carbon star and a very young main sequence star. However, we note
that the optical magnitudes of the carbon star components place them
at distances that are not compatible with SagDIG distance moduli. The
red star component could be a foreground star belonging to the
galactic population and therefore not necessarily carbon giant, but
also a carbon or M-dwarf.

Therefore we conclude that the solution with the two star SED points
towards the presence of a blend rather than a binary star. For two
objects there is also a possibility that they are high redshift
($z\sim 2.5$) galaxies. A conclusive test would need a spectroscopic
analysis of the three objects, which is clearly beyond the scope of
the present paper.
\end{appendix}

\end{document}